\documentclass[aip,jcp,showkeys,floatfix,12pt]{revtex4-1}
\usepackage{float}
\usepackage{graphicx}
\usepackage{amsmath}
\usepackage[usenames,dvipsnames]{xcolor}
\usepackage{amsfonts}
\usepackage{amssymb}
\usepackage[utf8]{inputenc}

\begin{document}

\title {Conformational degrees of freedom and stability of 
splay-bend ordering in the limit of a very strong planar anchoring}
\author{Lech Longa}
\author{Micha\l{} Cie\'sla}
\affiliation{Marian Smoluchowski Institute of Physics, Department of Statistical 
Physics and Mark Kac Center for Complex Systems Research,  Jagiellonian University, 
ul. \L{}ojasiewicza 11, 30-348 Krak\'{o}w, Poland.}
\author{Pawe\l{} Karbowniczek}
\author{Agnieszka Chrzanowska}
\affiliation{Faculty of Materials Engineering and Physics, Cracow University of Technology, 
ul.\ Podchor\c{a}\.{z}ych 1, 30-084, Krak\'{o}w, Poland.}
\date{\today}

\begin{abstract}
We study the self-organization of flexible planar trimer particles on a structureless
surface.  The molecules are made up of two mesogenic units linked by a spacer, all 
of which are modeled as hard needles of the same length. Each molecule can dynamically adopt two
conformational states: an achiral bent-shaped (\emph{cis-}) and a chiral zigzag (\emph{trans-}) one.
Using constant pressure Monte Carlo simulations and Onsager-type density functional theory (DFT),  
we show that the system consisting of these molecules exhibits a rich spectrum of (quasi-)liquid 
crystalline phases. The most interesting observation is the identification of stable 
smectic splay-bend ($S_{SB}$) and chiral smectic A ($S_A^*$) phases.
The $S_{SB}$ phase is also stable 
in the limit, where only cis- conformers are allowed. The second phase occupying 
a considerable portion 
of the phase diagram is $S_A^*$ with chiral 
layers, where the chirality of the neighboring layers 
is of opposite sign. 
The study of the average fractions of the \textit{trans-} and \textit{cis-} 
conformers in various phases 
shows that while in the isotropic phase all fractions are equally populated,
the $S_A^*$ phase is dominated by chiral conformers (zigzag), 
but the achiral conformers win in the smectic splay-bend phase. To clarify the possibility 
of stabilization of the nematic splay bend ($N_{SB}$) phase for trimers, 
the free energy of the $N_{SB}$ and $S_{SB}$ phases is calculated within 
DFT for the \textit{cis-} conformers, for densities where simulations 
show stable $S_{SB}$. It turns out that the $N_{SB}$ phase is unstable away from the 
phase transition to the nematic phase, and its 
free energy is always higher than that of $S_{SB}$, down to the transition to the nematic phase, 
although the difference in free energies becomes extremely small when approaching the transition. 
\end{abstract}

\maketitle

\section{Introduction}

\begin{figure}[htb!]
\centering
\includegraphics[width=0.6\columnwidth]{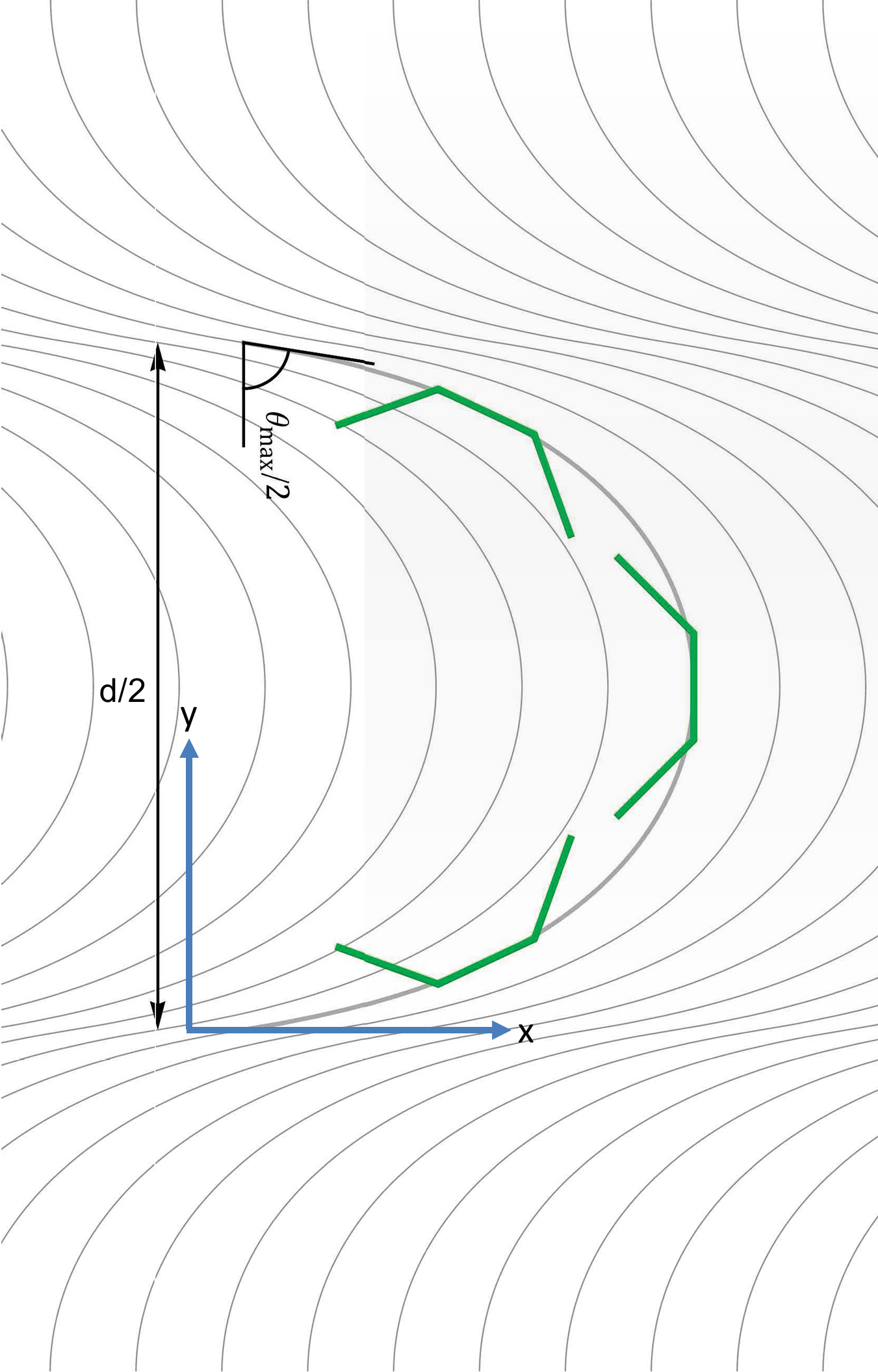}
\caption{Schematic representation of the periodic splay-bend modulation of the director
in $N_{SB}$ and $S_{SB}$. The local
director is tangent to the lines (gray) shown and
rotates between $\theta_{max}/2 $ and $-\theta_{max}/2 $ for $0\le y <d/2$, 
where $d$ is the period of the structure. 
Green lines show schematically bent-core molecular conformations 
that are responsible for the stabilization of the splay-bend ordering.
}
\label{paramnsb}
\end{figure}

Due to their unique shape anisotropy, achiral bent core 
liquid crystal mesogens can self-assembly in various mesophases, among which
the most unusual are the nematic twist-bend phase ($N_{TB}$) 
\cite{ref23,Cestari&DiezBerart2011,ref2,ref1,rewiew} and the
nematic splay-bend ($N_{SB}$) phase \cite{tavarone,karbowniczek,karbowniczek2}. 
In the $N_{TB}$ phase, there is no long-range positional ordering of the molecules
(as in ordinary nematics), but they follow a helix with the average local orientation
of the long molecular axis, the director $\mathbf{\hat{n}}(\mathbf{r})$, being
tilted with respect to the helical axis. The helical pitch of $N_{TB}$ is on the 10-nanometer 
scale, and the domains of the left and right induced twists are 
represented with equal probability. The experimentally observed first-order
phase transition from the nematic or isotropic phase to $N_{TB}$ 
is thus a rare example of \emph{spontaneous mirror symmetry breaking}.        
The $N_{SB}$ and the corresponding smectic splay-bend ($S_{SB}$) phases, 
which will be of our concern here 
are essentially 
an in-plane periodic splay-bend modulation of the director,
as shown schematically in Fig.~\ref{paramnsb}. In the $S_{SB}$ phase, there 
is also a quasi-long-range density modulation of molecular centers of mass that coincides
with that of the director. 
Modulated nematic structures have been predicted theoretically by  
Dozov  \cite{ref25} who pointed out that molecules 
whose predominant conformational states are bent ones, should  
tend to form a spontaneous local bend curvature 
of $\mathbf{\hat{n}}(\mathbf{r})$. 
However, unlike the ordinary cholesteric phase, where the three-dimensional (3D) space is
filled with a periodic and homogeneous twist \cite{ref36},  
a nematic state of pure spontaneous bend cannot be realized without
introducing defects, which are energetically expensive.
A less costly, defect-free, periodically modulated bend deformation is also
allowed if combined with some twist or splay \cite{ref23,ref25}. The
relative stability of these different variants is controlled by the ratio of 
the splay elastic constant to the twist elastic constant in the underlying 
nematic phase \cite{ref25,RevModPhys.90.045004}. 
Uniform periodic bend and twist, where the director simultaneously bends
and rotates in the cone, gives the aforementioned $N_{TB}$ phase.
Both $N_{TB}$ and $N_{SB}$ must be locally polar by symmetry, and, as it turns 
out, they do not exhaust the possibilities for stable one-dimensional periodic
nematic structures with nonzero bend \cite{ref48,ref22,extfield}. 

The family of mesogens known to exhibit $N_{TB}$ is quite  
substantial, obeying (most frequently) chemically achiral dimers 
\cite{Sepelj&Lesac2007,Panov&Nagaraj2010,Cestari&DiezBerart2011,%
Borshch&Kim2013,Chen&Porada2013,%
Paterson&Gao2016,Lopez&RoblesHernandez2016},  
bent--core mesogens \cite{Gortz&Southern2009,Chen&Nakata2014},
achiral oligomers \cite{trimer1,trimer2,oligomer}, or even 
the recent example of a
hydrogen-bonded complex \cite{hydrogen1}. In contrast, 
the $N_{SB}$ phase is scarce. 
It has been observed in colloidal bananas \cite{doi:10.1126/science.abb4536}, 
but not, to our knowledge, in bulk thermotropic materials. However, a transition 
$N_{TB}$ to $N_{SB}$
can be induced by an applied electric 
field \cite{extfield,PhysRevE.98.022704,doi:10.1126/sciadv.abb8212},
or as an interface between two homochiral $N_{TB}$ domains of opposite chiralities \cite{ref10}.
 It is not clear why the bulk $N_{SB}$ has not been detected so far, given the hundreds of
mesogens with $N_{TB}$. According to Dozov's theory\cite{ref25} the ordering $N_{SB}$ 
can spontaneously form if the splay elastic constant is less than twice 
the twist elastic constant, which seems to be in favor of at least some of the  
experimental data \cite{elastic1,elastic2}.  
Recently, using Monte Carlo simulations and mesoscopic modeling, it has been
demonstrated that the limitations of stabilizing $N_{SB}$  
can be circumvented in some cases if we confine bent core molecules to a planar 
surface \cite{tavarone,karbowniczek,karbowniczek2}. Here, the $N_{SB}$ phase
can win over $N_{TB}$ because development of twist deformations, which require   
a kind of an "escape into 3D", is hampered by the imposed constraints.

Clearly, none of the molecules that self-organize in $N_{TB}$ has a fixed geometry, and
the population of conformational molecular states, which is governed by a 
dynamic equilibrium, can be important (\emph{see e.g.} \cite{Archbold2017}
and references therein). Here, we study this effect on the formation of a stable
 splay-bend, nematic, and smectic orderings. Since, to date, the most probable scenario for obtaining 
stable splay-bend modulation is spatial confinement \cite{ref10}, or an external field 
\cite{extfield,PhysRevE.98.022704,doi:10.1126/sciadv.abb8212} we limit ourselves 
to a monolayer composed of molecules with conformational degrees of freedom. 

We generalize our two-segment hard V-shaped molecules \cite{karbowniczek,karbowniczek2}
to three segments of equal length (trimers), where the side ones are allowed 
to occupy different conformational states. Even in this simple model, the 
inclusion of conformations can be realized in many different ways 
(\emph{see e.g.} \cite{doi:10.1080/15421400701738586}). Here, we limit ourselves 
to the simplest possibility, where for each equivalent side segment of the molecule 
only two conformational states are allowed, as shown in Fig.~\ref{fig1}.
This leads to four conformations per molecule, where
the non-chiral bent ones (\emph{cis+},\emph{cis-}), Fig.~\ref{fig1} (b, c), are expected to support
stable splay-bend deformations 
\cite{tavarone,karbowniczek,karbowniczek2}, while the chiral (\emph{zigzag-}, \emph{zigzag+}) ones, 
Fig.~\ref{fig1} (a,d), are in favor of mirror symmetry breaking.
In addition to the main
aim of this work related to stabilization of splay-bend and related orderings, 
taking chiral conformers into account can contribute to our understanding of 
the problem of spontaneous mirror symmetry breaking in monolayers 
composed of flexible achiral mesogens,
especially whether it
is driven by flexopolarization \cite{ref23,ref25,Vaupotic2014} or some other
mechanism \cite{ref35,Goodby}. These studies should also be of importance 
in modeling the trans-cis isomerization that occurs when photo-switchable 
molecules are attached to a surface 
\cite{doi:10.1021/cr980078m,KATSONIS2007407,doi:10.1146/annurev.physchem.040808.090423,doi:10.1021/la102788j,cryst11121560}

We should add that, so far, \emph{zigzag+} and \emph{cis+} trimer particles 
subjected to planar confinement have been considered separately \emph{i.e.} without 
conformational degrees of freedom. More specifically, for purely \emph{zigzag+} trimer molecules, 
Pe'on \textit{et al.} \cite{peonek} showed the possibility of
stable nematic and smectic phases. The Onsager-type theory for this system was 
later developed by Varga \textit{et al.}
\cite{varga}, leading to similar predictions for liquid crystalline structures. 
By systematic manipulation of the shape anisotropy, they found that the \emph{zigzag-} 
shape with increasing bending angle and length of the side molecular segments favors 
the smectic order relative to the nematic one. 
Interestingly, in the smectic layers, the central segment of the trimer \emph{zigzag+}
was tilted with respect to the layer normal, and each layer behaved 
like an ideal gas of trimers (on average) with parallel orientation.

The study of \textit{cis+} trimers in 2D was initially carried out by Martinez \textit{et 
al.} \cite{martinez}. 
The authors found a quasi-long-range nematic and tetratic order, a curly nematic 
pattern similar to that of $N_{SB}$ (or $S_{SB}$), and the formation of chiral clusters despite 
the fact that the particles were achiral. All these works have been refined next by 
Tavarone \textit{et al.} \cite{tavarone}, where 
the authors quantified the difference in the thermodynamic behavior of the two 
molecular geometries. 
In addition to the Monte Carlo simulations in different ensembles, they 
implemented a more advanced Monte Carlo algorithm involving cluster moves to find 
evidence for the splay-bend deformations and 
to demonstrate that the isotropic quasi-nematic transition observed follows 
a Kosterlitz-Thouless disclination unbinding scenario.
They also used a simplified version of Onsager's Density Functional Theory to 
explain some of their results, such as the observed quasi-nematic--quasi-smectic 
phase transition. 

The model presented here that combines the properties of both types of particles 
is a generalization of the studies mentioned above.
Although this system seems interesting in itself, our choice is 
also motivated by the recent work of Jansze \emph{et. al.}\cite{hydrogen1}  
on $N_{TB}$ driven by hydrogen bonding in systems based on benzoic acid.
The authors attribute the stabilization of the $N_{TB}$ phase to the formation
of cyclic hydrogen-bonded supramolecular trimers,  similar to our bent (\emph{cis})
conformers. These trimers coexist in dynamical equilibrium with 
open hydrogen-bonded complexes, which, in turn, are similar to chiral (\emph{zigzag}) conformers 
and monomers (which are not present in our simplistic model). 

An interesting question then is whether our simplified
model, which also entails dynamical equilibrium \textit{trans-cis}
subjected to confined geometry,
can stabilize the splay-bend ordering (in analogy to the one observed 
in \cite{ref10}) and perhaps some other
quasi-liquid crystalline structures.
To examine such property-structure correlations,
we will carry out Monte Carlo simulations at constant pressure for a monolayer composed of 
$500 \le N \le 2000$  molecules shown in Fig.~\ref{fig1} along with an Onsager Density Functional analysis.  
The paper is organized as follows.
Section II defines the model that is next studied in Section III 
using Monte Carlo simulations at constant pressure.
The results of the simulations are further supported by 
predictions of Onsager's DFT formalism in 
Sections IV and V. The last section provides a brief summary 
along with the main conclusions.
\section{Model Particles}
\begin{figure}[htb]
\begin{center}
\includegraphics[width=0.5\columnwidth]{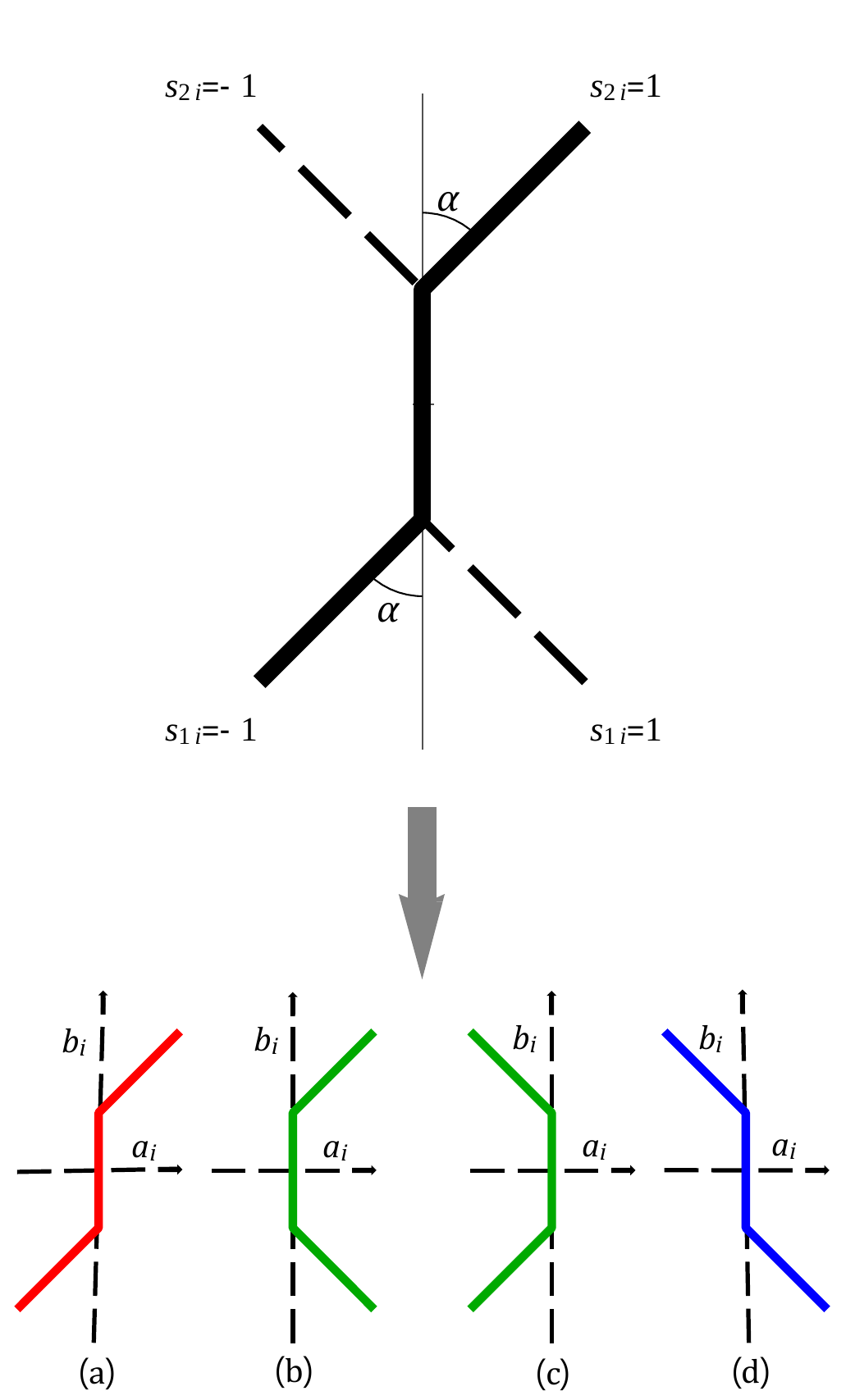}
\end{center}
\caption{Parametrization of particle's conformational degrees of freedom (top) 
and resulting four conformers (bottom). Particles are build out of three 
segments 
of the same length $l=1$. Each side segment can independently and with equal 
probability orient at angle $\pm \alpha$ with respect to the central segment, 
which we parameterize using two discrete variables  
${s_{ji}}=\pm 1; \,\, j=1,2$. 
Four resulting  conformers are:  bent (\emph{equiv.} \emph{ cis$\pm$})
conformations (b,c)  and  chiral \emph{zigzag$\mp$} (\emph{equiv.} 
\emph{trans$\mp$}) conformations (a,d). 
Note that (b) and (c) differ by the $\pi$-rotation while (a)  and 
(d)  are of opposite chirality in 2D. 
In simulations and DFT calculations we use local, molecule-attached orthogonal 
system with $b_i$ axis kept parallel to the central 
molecular segment and $a_i$ axis perpendicular to it. 
The  $\{a_i,b_i\}$ axes cross in the mid-point 
of the central segment. 
}
\label{fig1}
\end{figure}

We study a simple athermal model of hard flexible molecules whose positions
and orientations are limited to a structureless planar surface or, equivalently, to be
subjected to a strong planar anchoring. Each molecule is made up of three
segments of the same length $l=1$, with flexible
arms attached to the central one. Arms can adopt two 
conformational states: achiral bent (\textit{cis$\pm$} ) and chiral
(\emph{zigzag$\pm$})
parameterized by dynamic variables $s_{\beta i}=\pm 1$, with $\beta$ 
labeling
the arms of the $i-th$ molecule (Fig.~\ref{fig1}). The relative 
orientation of the arm to the central segment is given 
by a single angle $\alpha$. In the limiting case of $\alpha=0$, the 
molecule is reduced to a needle of length 3. 
The relative population of the conformes 
is governed by purely excluded volume interactions and a given fixed pressure 
or density.

As mentioned above, monolayer studies to date indicate that bent 
molecular architectures should support stable splay-bend-type deformation 
\cite{tavarone,karbowniczek,karbowniczek2}. Therefore, adding conformations that 
preserve bent is not expected to qualitatively change this picture, 
and in our elementary model, we represent all possible bent conformations
(\emph{see} Fig.~\ref{fig1}) by a single (averaged) bent shape. 
The addition of \emph{zigzag} conformations of opposite chiralities should 
favor smectic ordering and mirror symmetry breaking. However, it is of interest to
observe the entropic competition between these two effects on the phase sequence.
To identify stable structures and their properties and locate phase boundaries,
we performed NPT Monte Carlo simulations at a constant number of particles (N), 
constant pressure (P), 
and constant temperature (T), supported by Onsager density functional theory 
and bifurcation analysis. 
We limit ourselves to sampling molecular translational, orientational, and conformational states. 
The kinetic energy part of the partition function is integrated over momenta. 
For the $i-th$ particle, 
it leads to a factor, say $\Lambda({s_{1i},s_{2i}},T)$, which depends on the conformational 
degrees of freedom $s_{\alpha i}$  \cite{doi:10.1080/15421406.2011.572014}. 
To keep our model as simple as possible and to avoid 
parameters that are difficult to control, we disregarded the 
conformers' energy landscape. In the
same spirit, we replace $\Lambda({s_{1i},s_{2i}},T)$ by its 
average value $\overline{\Lambda(T)}$, 
where the average is taken over the degrees of freedom of the molecular conformations. 
With this assumption, 
the average number of conformers  
in a given structure is governed exclusively by the packing entropy. Therefore, in the
reference isotropic phase, at least at low densities, their average fractions should be the same.
\section{Monte Carlo results}
\label{sectionMC} 
 Monte Carlo simulations were performed in an NPT ensemble. 
The system consisted of particles $N=500$ to $N=2000$ 
in a square box with side length $L$ and periodic boundary conditions applied. 
In the MC cycle, we probed a new configuration for randomly chosen N molecules. 
A trial molecular configuration was generated by a random move of the center of mass 
of the molecule, a random
rotation of the local molecular frame and with the probability of $0.1$ a random 
change in molecular shape by choosing one of the four possible conformations
(\emph{see} Fig. \ref{fig1}) 
with the same probability. The step was accepted if the particle did not intersect with the others. 
After every ten cycles, the size of the box was adjusted to keep the pressure constant. 
Rescaling kept the box square.
In simulations 40\% to 50\% of new molecular configurations, system rescaling and 
conformational changes
were accepted. Lower acceptance ratios were used for dense and highly ordered states, 
allowing the solution to partially overcome metastability traps.
The reduced density was calculated as
\begin{equation}\label{redDens}
 \bar{\rho}=N l^2/S \equiv N/S,   
\end{equation}
where $S=L^2$ is the surface area and the length $l$ of 
the molecular segment is $l=1$. 
Each simulation was initialized from a random, disordered gas of 
conformers within a large box, and the pressure was increased in
adjustable steps until the transition to (quasi-)liquid crystalline phases occurred. 
The equilibration appeared slow and took about $2.5 \cdot 10^6$ cycles, 
after which the data were collected for the analysis of the structure once per 10
cycles in a production run of $10^4$ cycles. To verify that the equilibrated 
configurations were not just metastable states, we used different initial states, 
including perfectly oriented polar and antipolar nematic/smectic configurations.

\begin{figure}[htb]
\begin{center}
\includegraphics[width=0.4\columnwidth]{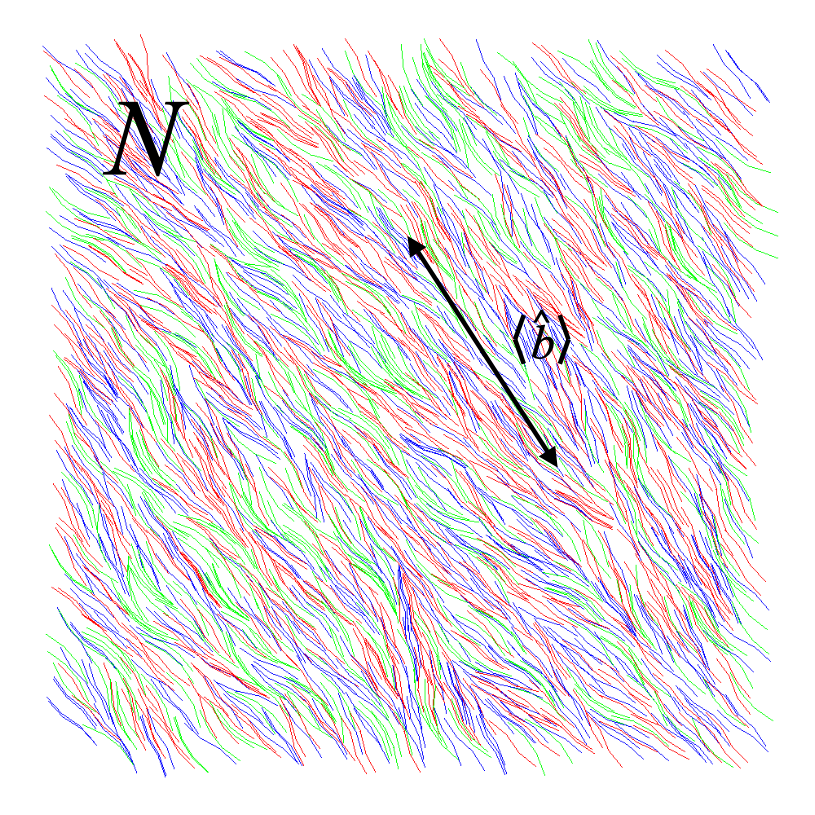}
\vspace{0.02\columnwidth}
\includegraphics[width=0.4\columnwidth]{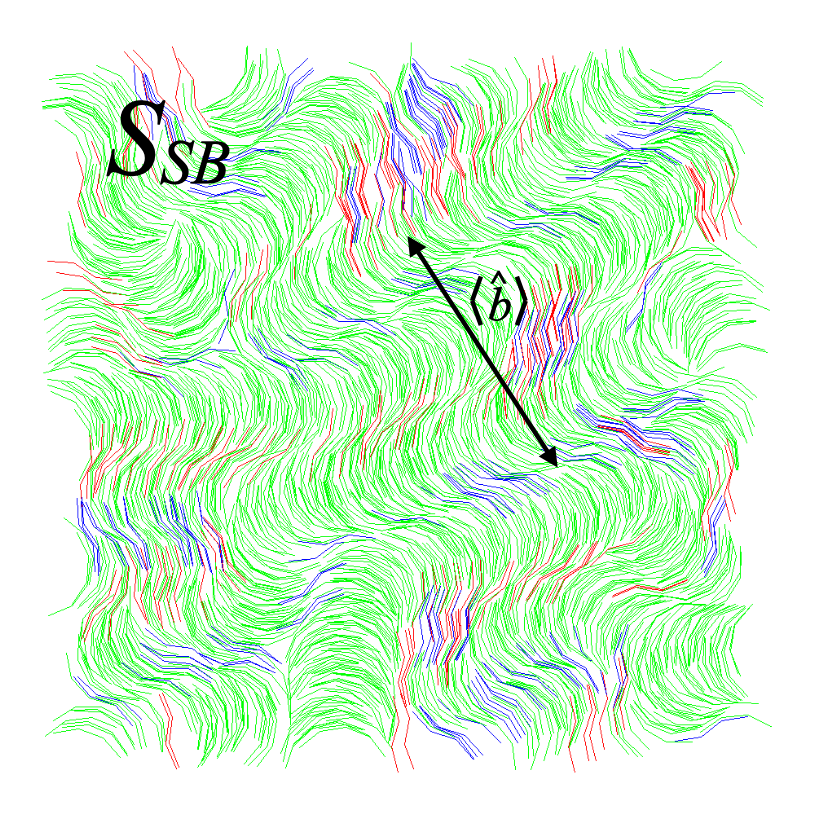}
\hspace{0.02\columnwidth}
\includegraphics[width=0.4\columnwidth]{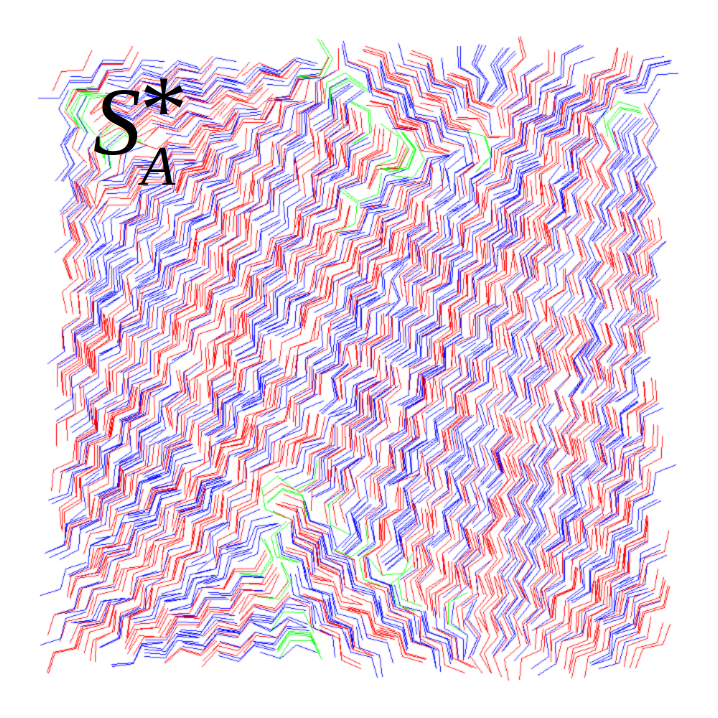}
\vspace{0.02\columnwidth}
\includegraphics[width=0.4\columnwidth]{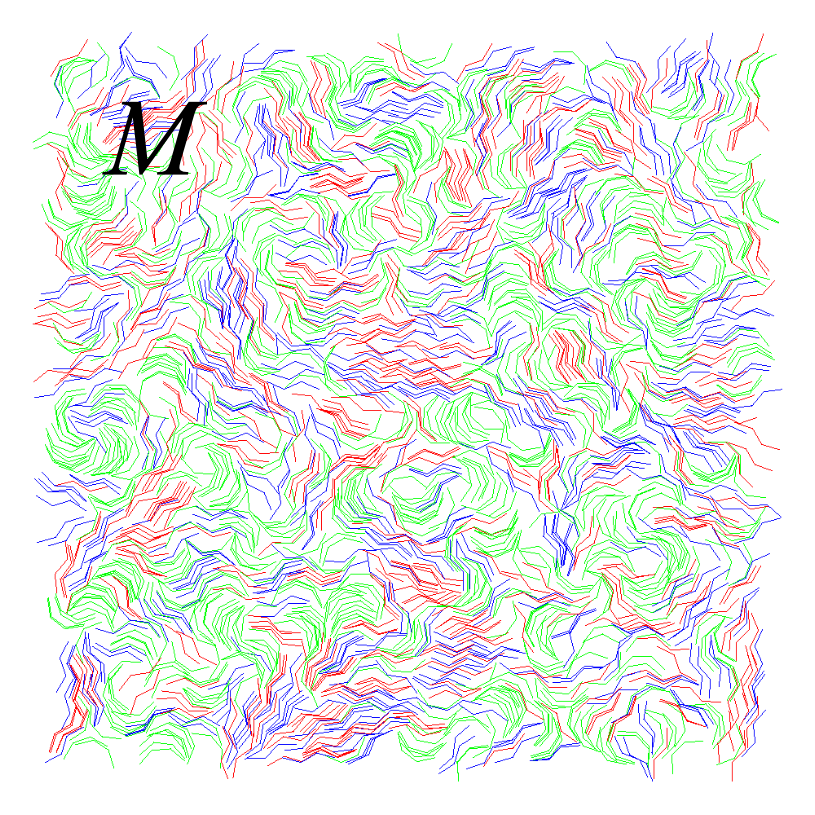}
\end{center}
\caption{Examples of ordered structures found in simulations: 
nematic ($N$) for $\alpha=\pi/12$ and  $\bar{\rho}=2.0$, smectic splay-bend ($S_{SB}$) for 
$\alpha=\pi/6$ and $\bar{\rho}=3.1$, 
and {(antichiral)} smectic A ($S_{A}^*$) for $\alpha=\pi/3$ and 
$\bar{\rho}=2.5$; $\langle \mathbf{\hat{b}}
\rangle= \langle \frac{1}{N} \sum_{i=1}^N b_i \rangle$
stands for the average orientation of the central molecular segments.
The last picture presents a mix ($M$) of differently ordered 
domains ($\alpha=\pi/4$, $\bar{\rho}=1.39$).
In isotropic and nematic phases fractions of red and blue conformers (Fig. \ref{fig1})
are statistically the same. In $S_{SB}$ achiral conformers dominate, and in $S_{A}^*$ they disappear. 
Color coding is used to distinguish between different conformers 
(see caption to Fig.\ref{fig1}). }
\label{fig:confy}
\end{figure}

Interestingly, we observed five distinct liquid phases: isotropic ($I$), nematic ($N$), $S_{SB}$, 
smectic A with chiral layers of alternating chirality ($S_{A}^*$), in which both types of particles 
are tilted relative to each other, and coexisting domains with different local order, 
which we denoted as $M$. Typical configurations of ordered 
structures are presented in Fig. \ref{fig:confy}. Color coding is used for different conformers:
red and blue for chiral \textit{zigzag} and green for achiral bent core ones
(\textit{cis}), respectively.

\begin{figure}[htb]
\begin{center}
\includegraphics[width=0.45\columnwidth]{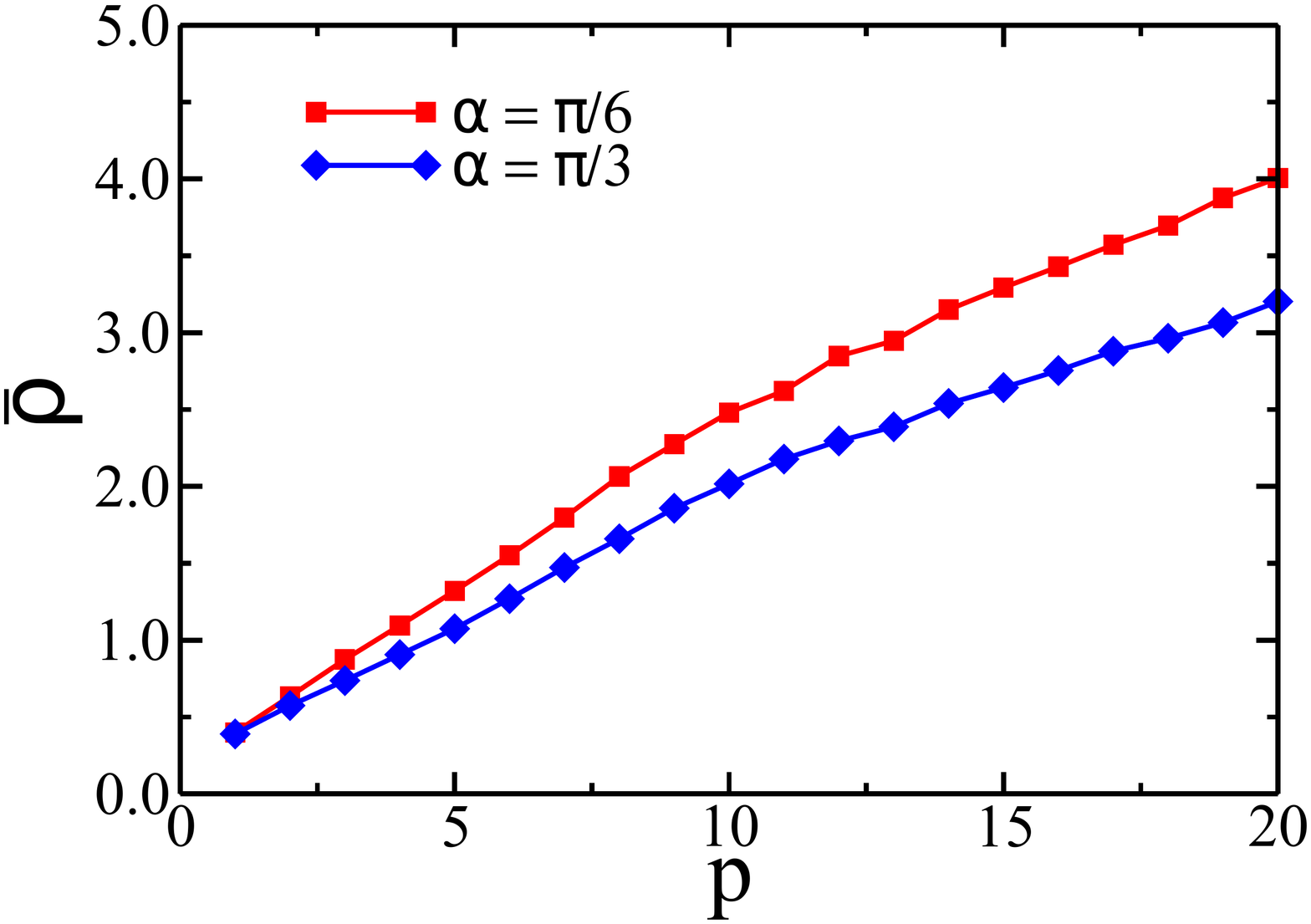}
\hspace{0.05\columnwidth}
\includegraphics[width=0.45\columnwidth]{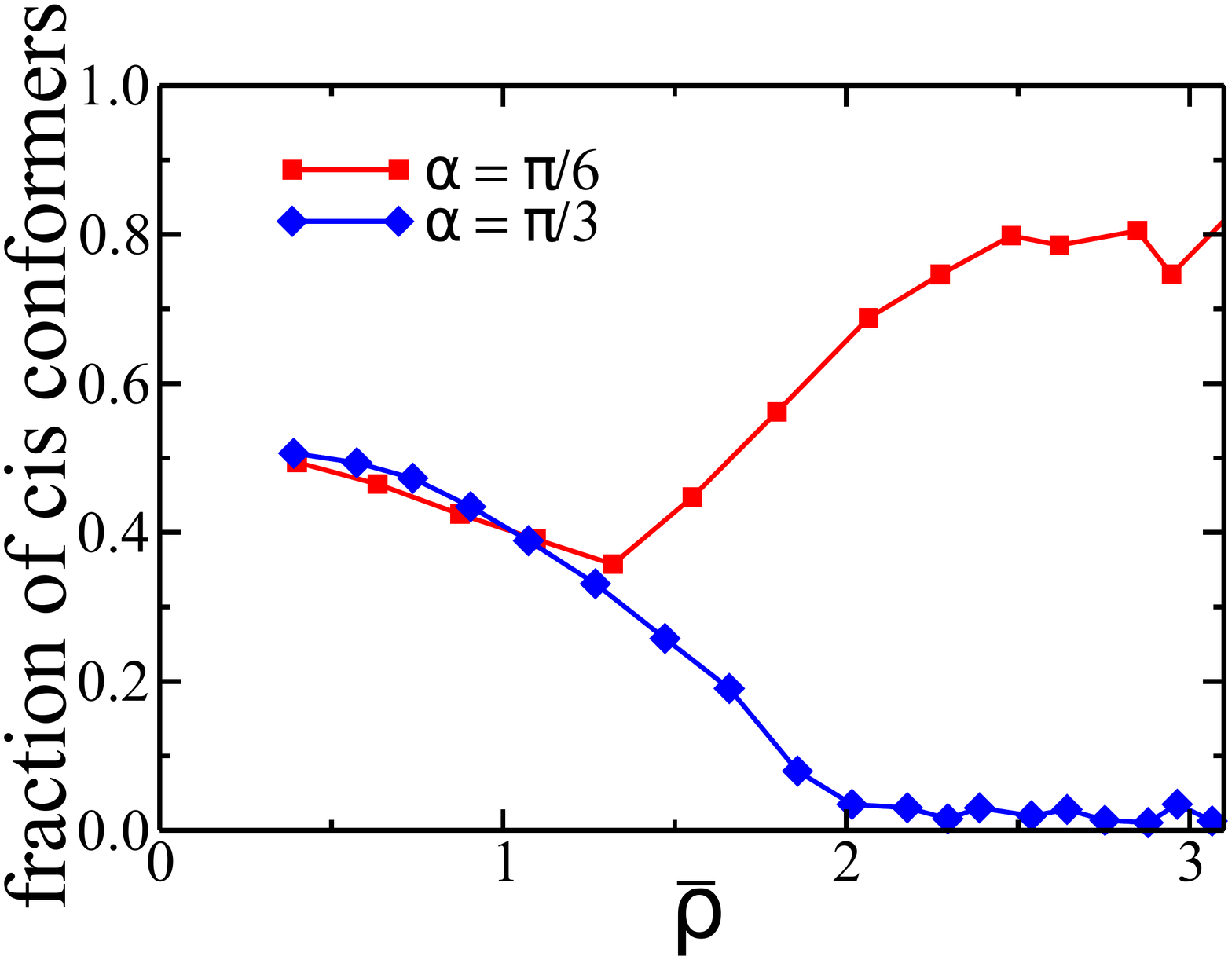}
\end{center}
\caption{Dependence of density on pressure (left) and fraction of \textit{cis}
conformers (the green one in Fig.~\ref{fig:confy}) on density (right)
for $\alpha=\pi/6$ and  $\alpha=\pi/3$.}
\label{fig:eosandconfs}
\end{figure}

\begin{figure}[htb]
\begin{center}
\includegraphics[width=0.45\columnwidth]{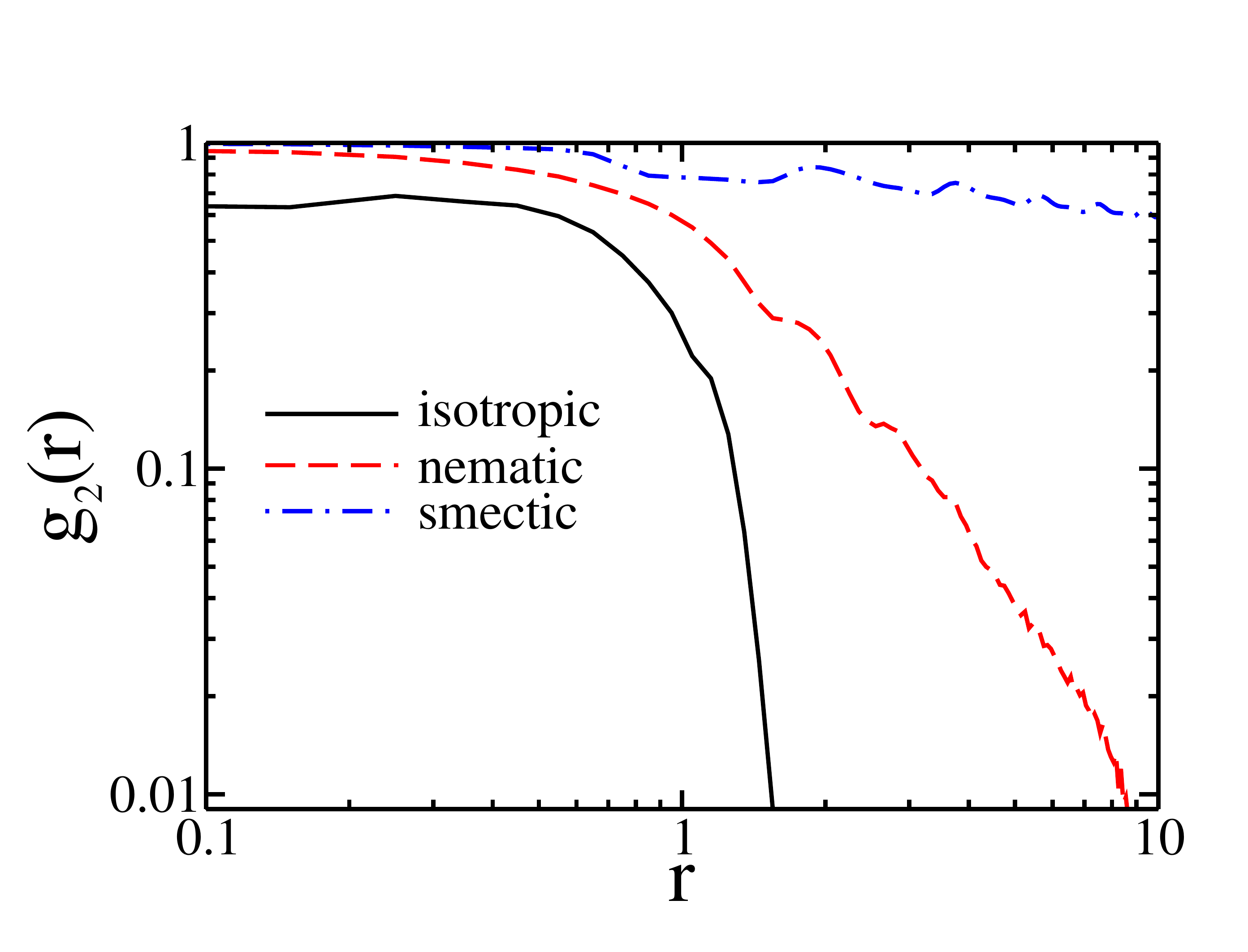}
\hspace{0.05\columnwidth}
\includegraphics[width=0.45\columnwidth]{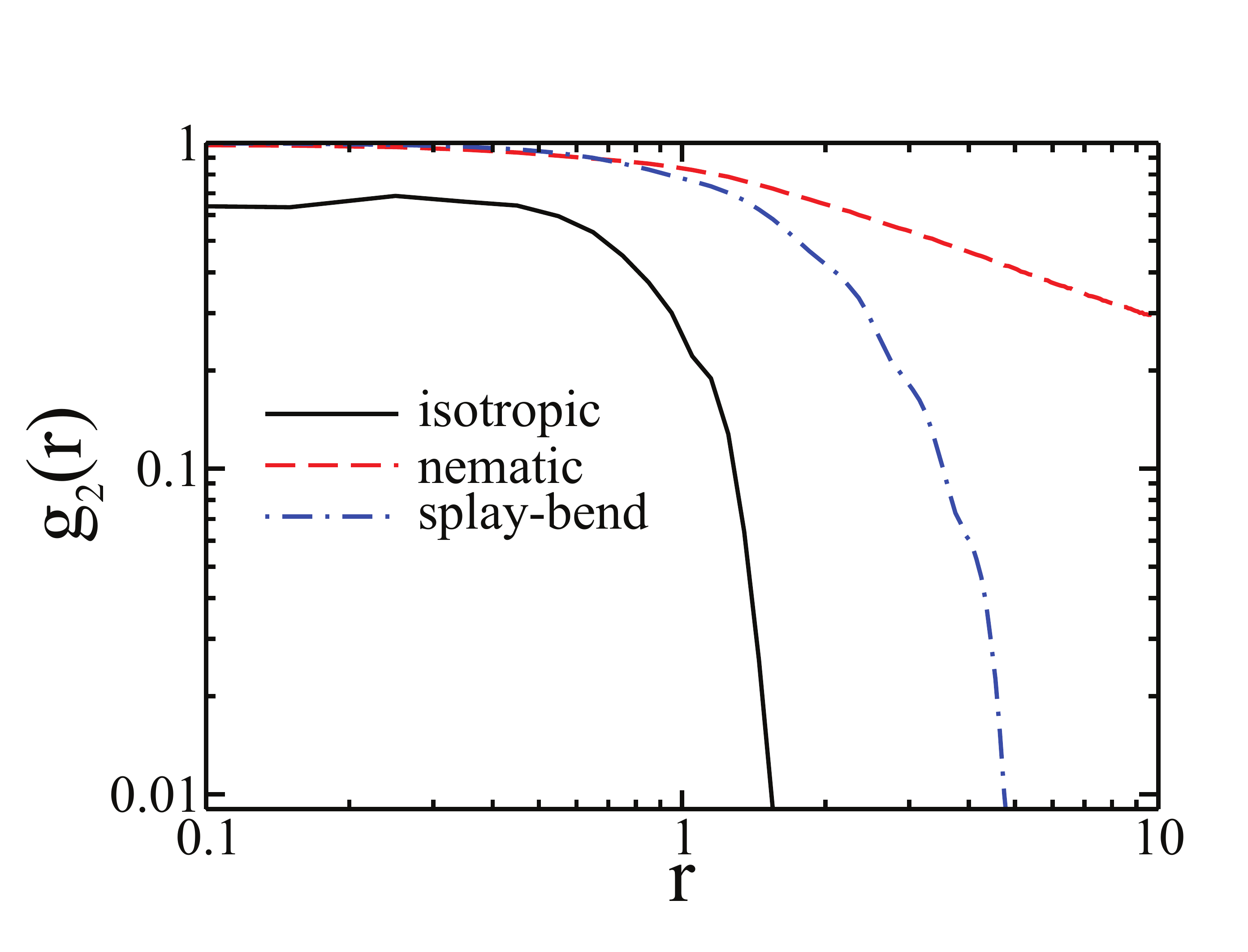}
\end{center}
\caption{Examples of bond-order correlations. The plot on the left corresponds 
to $\alpha=60^{\circ}$. The solid black line was obtained for $\bar{\rho} = 0.4$ 
and corresponds to the isotropic phase. The dashed red line is for $\bar{\rho} = 1.5$ and 
is typical for nematics, while the blue dashed line ($\bar{\rho} = 2.5$) shows quasi-smectic 
ordering. For $\alpha=30^{\circ}$ (right panel), the red dashed line ($\bar{\rho} = 1.6$) 
corresponds to the nematic phase and the blue dashed line ($\bar{\rho} = 3.1$) corresponds 
to the smectic splay-bend order.}
\label{g2corr}
\end{figure}

The fractions of different conformers change from phase to phase. 
In general, the \textit{cis} conformers disappear with increasing density. 
In a well-established $S_{A}^*$ phase, where the average main axis of the 
inertia tensor of the \textit{zigzag} 
conformers remains parallel to the layer normal,
there are almost no achiral \textit{cis} conformers. 
The only exception to this behavior was found in 
the $S_{SB}$ phase. In that case, a small fraction of the \textit{zigzag} conformers is 
located between the splay-bend layers
(\emph{see} the upper right panel of Fig. \ref{fig:confy}). 
The dependence of the fractions of the achiral \textit{cis} conformers on 
the packing density for the transitions $I-S_{SB}$  
and $I-S_{A}^*$ is shown in Fig. \ref{fig:eosandconfs}.
Note that there are no discontinuities in this dependence and 
in the dependence of the density on the pressure.
Therefore, to quantitatively distinguish the observed structures, 
we use the bond correlation function $g_2(r)$:
\begin{equation}
g_2(r)=\left \langle \frac{\sum_{i=1}^{N}\sum_{j=i+1}^{N}P_2\left 
( \boldsymbol{u}_i \cdot \boldsymbol{u}_j \right )\delta \left ( r-r_{ij} \right ) 
}{\sum_{i=1}^{N}\sum_{j=i+1}^{N} \delta \left ( r-r_{ij} \right )} \right \rangle
\label{rownanie22}
\end{equation}
since it is a well-established method of phase differentiation (see, for example, 
\cite{tavarone,cinacchi,schlotthauer}). Exemplary correlations $g_2(r)$ are presented in 
Fig. \ref{g2corr}. The transition from $I$ to $N$ and the subsequent transition to smectics 
occur along with the increase in the correlation length. In the smectic phases, 
there are also visible peaks in the $g_2(r)$ function that indicate density modulation 
along the propagation of the wave vector. However, when the transition to $S_{SB}$ occurs, 
there is a significant reduction in the correlation length, due to the period of 
the splay-bend slabs formed.

\begin{figure}[htb]
 \begin{center}
 \includegraphics[width=0.9\columnwidth]{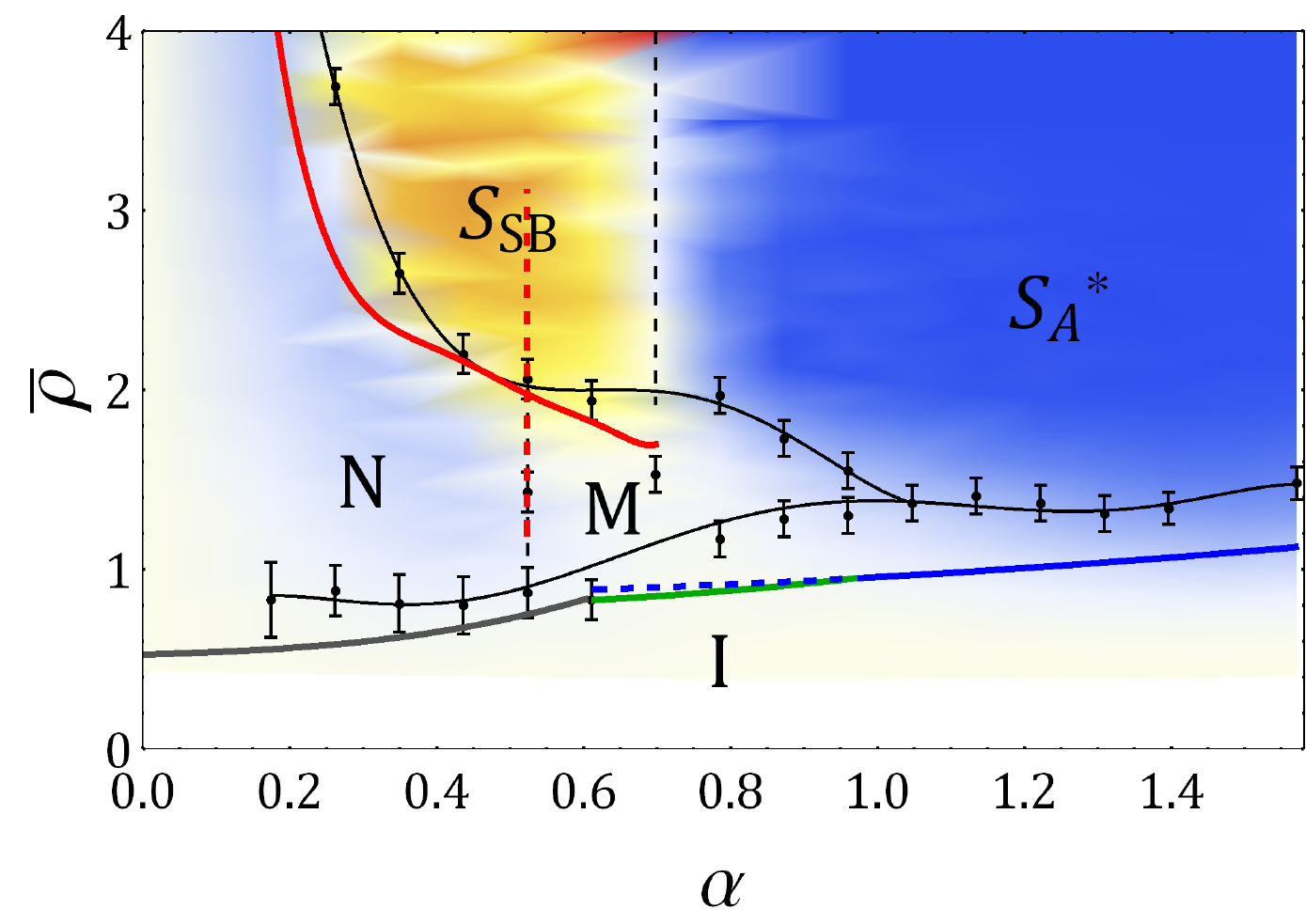}
\hspace*{1.6cm}
\includegraphics[width=0.6\columnwidth]{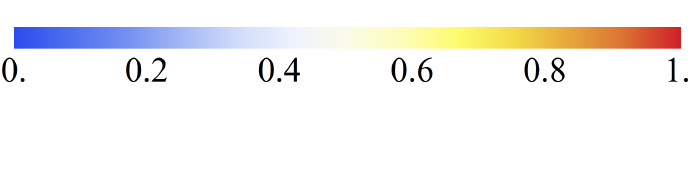}
 \end{center}
 \caption{Phase diagram obtained from Monte Carlo simulations  and DFT
 analysis. Background colorization, corresponding to the fraction of \textit{cis} conformers, 
is explained in the horizontal legend. Blue color corresponds to the domination of 
 \textit{trans} particles, whereas red/yellow for $cis$ ones. 
 Thin lines with error bars correspond to data from MC simulations.
 Solid thick lines 
are obtained from Onsager's DFT analysis as explained in the body of the paper. 
These lines correspond to the transitions from the isotropic phase to nematic 
(dark grey), mixed phase (green) and smectic A (blue line) phases using 
analysis with full orientational degrees of freedom. The blue line is 
extended to the mixed phase region and presented as a dashed line in order 
to help to explain complicated behaviour in this phase further in the text. 
Thick red line corresponds to the transition from ideally oriented nematic 
to splay-bend phase. Along dashed red line a full numerical solution of DFT 
is obtained for \textit{cis} particles
(\emph{see} Section IV). }
 \label{fig:phase_diagram}
 \end{figure}

Using $g_2(r)$ and abrupt changes in the average conformer population as a transition indicator, 
we constructed the phase diagram of our system, Fig. \ref{fig:phase_diagram}. 
We observed $I-N-S_{SB}$ and 
direct $I-S_{A}^*$ phase transitions as a function of density. Furthermore, 
the coexisting domains of $N$, $S_{SB}$, and $S_{A}^*$
were identified and denoted $M$. This mix of phases appeared independently 
in the starting configuration for the MC simulations; we tried all: 
$I$, $N$, $S_{SB}$, and $S_{A}^*$ as initial configurations, and in all cases 
we end up with the $M$ phase. Similarly, we checked the stability of the remaining phases.
As expected, for low densities, the isotropic phase is observed. 
For higher densities, the behavior of the system 
depends on the angle $\alpha$. When it is small, only the nematic phase stabilizes. 
For higher values of $\alpha$,
up to $\alpha < 2\pi/9$, the high-density phase is $S_{SB}$. When $\alpha > 2\pi/9$, 
it is replaced by $S_{A}^*$. 
The background color of the phase diagram corresponds to the fraction of \textit{cis} 
conformers. Bow-shaped 
conformers dominate only in the $S_{SB}$ phase. In the $N$ and $S_{A}^*$ phases, this 
type of conformer
disappears with density growth.

\begin{figure}[htb]
\begin{center}
\includegraphics[width=0.75\columnwidth]{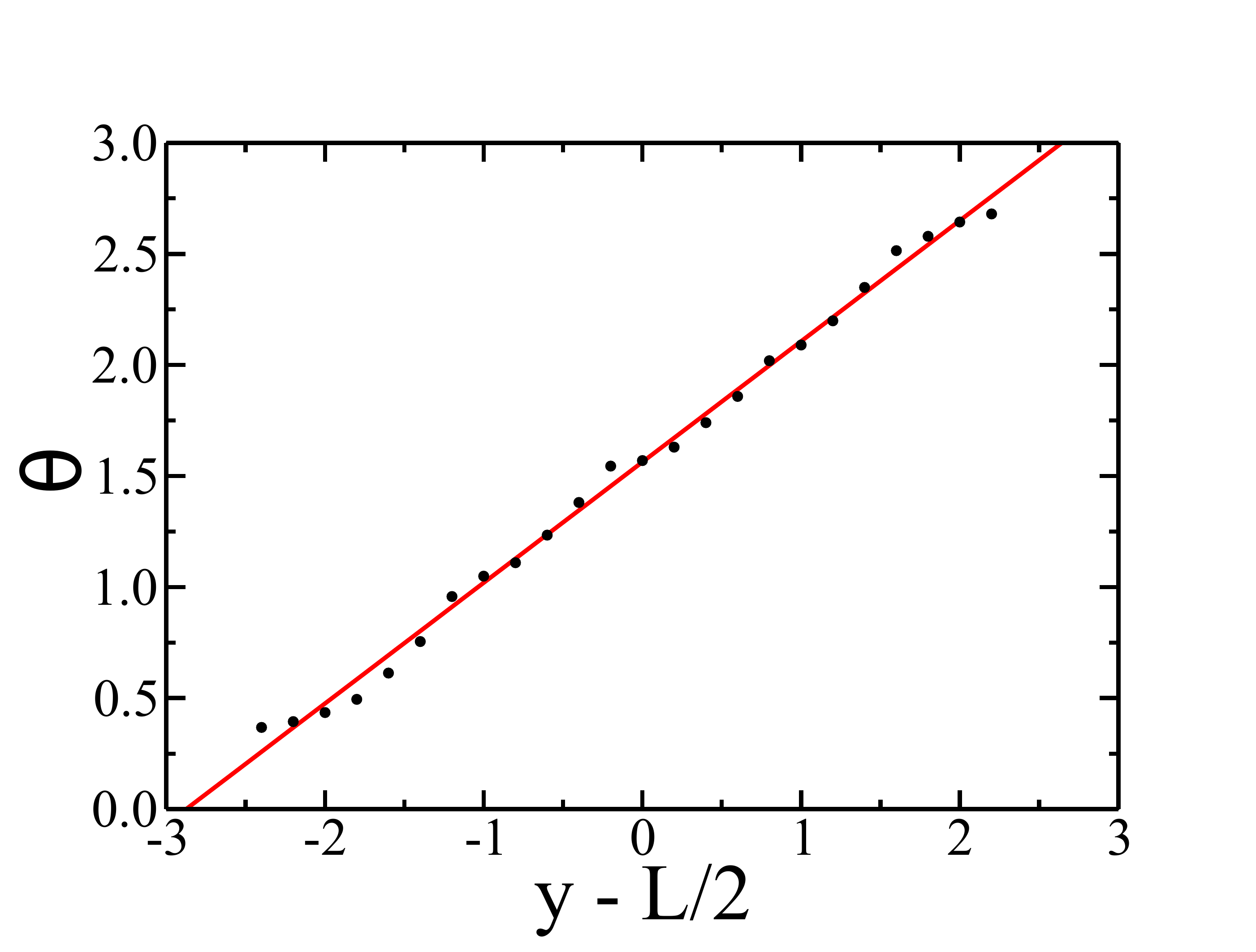}
\end{center}
\caption{Dependence of the mean particle orientation on its distance 
from center of a splay-bent layer. 
Results were calculated for a single layer taken from the system described by $\alpha=\pi/6$ 
and $\bar{\rho} \approx 3.0$. The black dots are the measured data and the red solid 
line is the linear fit: $\theta = \pi/2 + 0.54353 \cdot (y-L/2)$, where $y-L/2$ is 
the distance from the layer's center.}
\label{fig:order_profile}
\end{figure}

Finally, we also checked the profile of director's orientation $\theta(y)$ 
in the splay-bent layer. This dependence has to be assumed prior to 
any analytic calculations based on density functional theory. 
In previous studies, several different profiles were used. To obtain the profile, 
we extracted a single layer from the snapshot taken for 
$\alpha=\pi/6$ and $\bar{\rho} \approx 3.0$, where the $S_{SB}$ phase dominates. 
Then, the dependence of the average orientation of a particle on its distance 
from the center of the layer 
was measured. The results are shown in Fig.\ref{fig:order_profile}.
For the width of the layer here to be slightly less than $5.0$,
the director orientation profile is fairly well 
approximated by the linear function. Furthermore, 
the direction of the director in a single layer does not cover the full possible spectrum 
of angles $(0, \pi]$, but is limited to $[0.15 \pi, 0.85 \pi]$. 
It indicates in a very high orientational order in $S_{SB}$,
with the director rotating between $\pm \theta_{max}/2$ (Fig. 1)
as we proceed along the $y-$ axis.
\section{Density Functional analysis}
Following our previous work \cite{karbowniczek}, to identify stable structures 
and their properties and locate phase boundaries,
we supplemented computer simulations with analysis based on
Onsager-like density functional theory (DFT). We start with a brief summary of 
the Onsager formalism for our model, \emph{see} Fig.\ref{fig1}.  
Neglecting the external field terms, 
the Onsager non-equilibrium free energy functional, $\mathcal{F[\rho]}$,
for a system of molecules with conformational degrees of freedom subjected to planar 
confinement is given by 
\begin{eqnarray}
{\beta\mathcal{F}\left[ \rho \right]}
&=&
~\underset{(X)}{Tr}\left\{ \rho (X)%
\left[\, \ln \left( \overline{\Lambda(T)} \rho (X)\right) -1\right] \right\} 
-\frac{1
}{2}\underset{(X_{1},X_{2})}{Tr}\left[ \rho (X_{1})f_{12}\,\rho
(X_{2})\right],   \label{freeenergy}
\end{eqnarray}
where
$f_{12}=e^{-\beta V(X_{1},X_{2})}-1 $ 
stands for the Mayer function, $V(X_{1},X_{2})$ is the interaction 
potential, $\beta=\frac{1}{k_BT}$ is
the reduced temperature with $k_B$ being the Boltzmann factor, and $\rho (X)$ is the
local particle density routinely normalized to the total number of
molecules:
\begin{equation}\label{normalization}
 {Tr}_{(X)}\rho(X)=N.    
\end{equation}
The variable $X_{i}\equiv \{ {\mathbf{r}_i},\theta_i, \mathbf{s}_i  \}$ 
($X\equiv \{ {\mathbf{r}},\theta, \mathbf{s}  \}$)
represents the translational, orientational, and conformational degrees of freedom 
of the molecule '$i$', where $\mathbf{r}_i=(x_i,y_i)$ is the position of the 
center of the central molecular segment of the particle, and 
$\mathbf{s}_i=(s_{1i},s_{2i})$ parameterizes the molecular conformational degrees of freedom 
with $s_{\beta i}=\pm 1$ and $\beta=1,2$. The molecular orientation, given by
the angle $\theta_i$, is measured
between the central molecular segment (parallel to the $b_i$ axis of the 
molecule-fixed frame $\{a_i,b_i\}$,
denoted by the dashed thin line in Fig.~\ref{fig1}) and the $y-axis$ of the 
laboratory reference frame and 
\begin{equation}
{Tr}_{(X_i)}=\sum_{s_{1i}=\pm1,\: s_{2i}=\pm1}^{ } \int_{0}^{L}dx_i\int_{0}^{L}dy_i \int_{-\pi}^{\pi}
d\theta_i.
\label{rownanie090}
\end{equation}
For hard-core interactions, the Mayer function is
negative of the excluded interval:
\begin{equation}
f_{12}=e^{-\beta V(X_{1},X_{2})}-1=-\Theta \left[ \xi ({\hat{\mathbf
r}}_{12},\theta_1,\theta_2,\mathbf{s}_1,\mathbf{s}_2)-  {r}_{12}\right], 
\label{f12teta}
\end{equation}
where ${\hat{\mathbf r}}_{12}=\frac{\mathbf{r}_{12}}{{r_{12}}}=
\frac{\mathbf{r}_{2}- \mathbf{r}_{1} }{|\mathbf{r}_{2}- \mathbf{r}_{1} |}$ and
$\xi$ is the contact function
defined as the
distance of contact from the origin of the second molecule for a given
direction ${\hat{\mathbf
r}}_{12}$, orientations $\theta_1, \theta_2$ and internal states $\mathbf{s}_1, \mathbf{s}_2$;
$\Theta$ denotes the Heaviside function. The contribution of kinetic energy is represented by 
$\overline{\Lambda(T)}$.

Now, by introducing the probability density distribution function $P(X)$ such that
$\rho (X)=NP(X)$ ($\underset{(X)}{{Tr}} P(X)=1$)
and ignoring the terms in (\ref{freeenergy}) that can be made independent of $P$, 
the reduced free energy per molecule, $f(P)$, can be written as
\begin{eqnarray}
f(P)=
\underset{(X)}{
{Tr}}\left[ P(X)\ln  P(X) \right] +\frac{
{\bar{\rho }}}{2}\,\underset{(X)}{
{Tr}}\left[ P(X)H_{eff}(X)\right],
\label{freeener}
\end{eqnarray}
where, as previously,  $\bar{\rho}$ is the reduced density (\ref{redDens}) 
and $H_{eff}$ is the effective excluded volume 
averaged over the probability distribution of particle "2" ($X_1 \equiv X$):
\begin{equation}
H_{eff}(X_{1})= S\underset{(X_{2})}{
{Tr}}\left\{
\Theta \left[ \xi (X_{1},X_{2})-r_{12}\right]\, P(X_{2}) \right\}.
\label{Heff13}
\end{equation}
The equilibrium distribution function is obtained by minimizing the functional 
of $f(P)$ over $P$, subject to the normalization condition. 
The resulting non-linear integral equation
for the stationary distribution $P(X)$ reads
\begin{equation}
P(X)=Z^{-1}e^{-\bar{\rho}\, H_{eff}(X)},
\label{self1}
\end{equation}
where the normalization constant $Z$ is given by
\begin{equation}
Z=\underset{(X)}{
{Tr}}\left[e^{-\bar{\rho}\,H_{eff}(X)}\right].
\label{self2}
\end{equation}

The symmetry of the pair interaction potential guarantees that the 
isotropic liquid state corresponding to $P(X)=P(s)$  always satisfies Eq.~(\ref{self1}).
Remaining solutions of Eq.~(\ref{self1}), where one-particle probability distribution
depends also on molecular orientations and/or positions. We will study these solutions
in the next part of the paper. For the analysis of liquid crystalline phases, 
we can omit structures 
with 2d spatial periodicity (solids). The structures left obey  
nematics, and smectics. 
They can be analyzed by solving the integral equation (\ref{self1})
with the ansatz 
\begin{equation}\label{ansatzForP}
P(X)=P(y,\theta,s).    
\end{equation}
\subsection{Which phase is stabilized: $N_{SB}$ or $S_{SB}$}
Although Monte Carlo results unequivocally recognize the stable phases for 
flexible trimers, in particular $S_{SB}$, 
it is of interest to confront simulations with the density functional analysis, 
Eqs.(\ref{freeener}-\ref{ansatzForP}).
Such studies can provide further insight into the problem of why
$N_{SB}$ is so scarce.  According to Anzivino, van Roij and Dijkstra 
\cite{doi:10.1063/5.0008936,PhysRevE.105.L022701}
the problem lies in the character of the splay-bend director modulation, 
which promotes a coupling to the one-dimensional density wave. This, in turn, 
leads to the splay-bend 
phase of smectic symmetry. The prediction was based on the grand-canonical 
Landau-deGennes theory. Here, we look at this problem 
from the perspective of Onsager DFT. As the $S_{SB}$ phase of flexible trimers 
is dominated by the \emph{cis} conformers (about 80\%), we started our DFT analysis
neglecting the \emph{zigzag} ones and assuming $s_{1i}=s_{2i}=1$ (Fig. 2). 
With this simplification and with periodic boundary conditions we performed  
free-energy minimization, Eq.~\ref{freeener}, 
along the dashed red line in Fig.~\ref{fig:phase_diagram} 
with full dependence on the spatial and angle variables of 
the distribution function (\ref{ansatzForP}). In the case of smectics, 
the size of the smectic layer has also been determined by the free energy minimization. 

The main equation (\ref{self1}) of DFT has been solved numerically in a self-consistent 
manner with the use of Gaussian quadrature for
the integrals. The distribution function (along with the excluded
volume kernel) has been approximated by 360 spatial points 
in the variable $y$ and 100 points in the angle $\theta$.
Details of this technique have previously been described 
in \cite{Agnesferro,AgnesSmek}.

Stationary solutions were obtained by starting the self-consistent 
calculations with an initial distribution of the following form:
\begin{eqnarray}
P(y,\theta)&=&Z^{-1}{\rm{exp}}\left( a_0 \,{\rm{cos}}(\theta) 
+ a_1 \,{\rm{cos}}(2\theta) + a_2 \,{\rm{cos}}(2 \pi y/d)
+  a_3\, {\rm{cos}}(4\pi y/d)\right. \nonumber \\
&+& \left. a_4\, {\rm{cos}}(2 \pi y/d)      
{\rm{cos}}(\theta)+a_5\, {\rm{cos}}(4 \pi y/d)      
{\rm{cos}}(2 \theta)\right),
\label{ini_condition}
\end{eqnarray}
where $a_i \,\,\, (i=1...5)$ are the numbers chosen at will from the interval $(-1,1)$ and
the initial value of $d$ was taken to be consistent with the simulations.
At high reduced densities, the structure stabilized by this procedure was $S_{SB}$, 
in agreement with the MC simulations. 

\begin{figure}[htb]
\begin{center}
\vspace{0.4cm}
\includegraphics[angle=90,width=0.666\columnwidth]{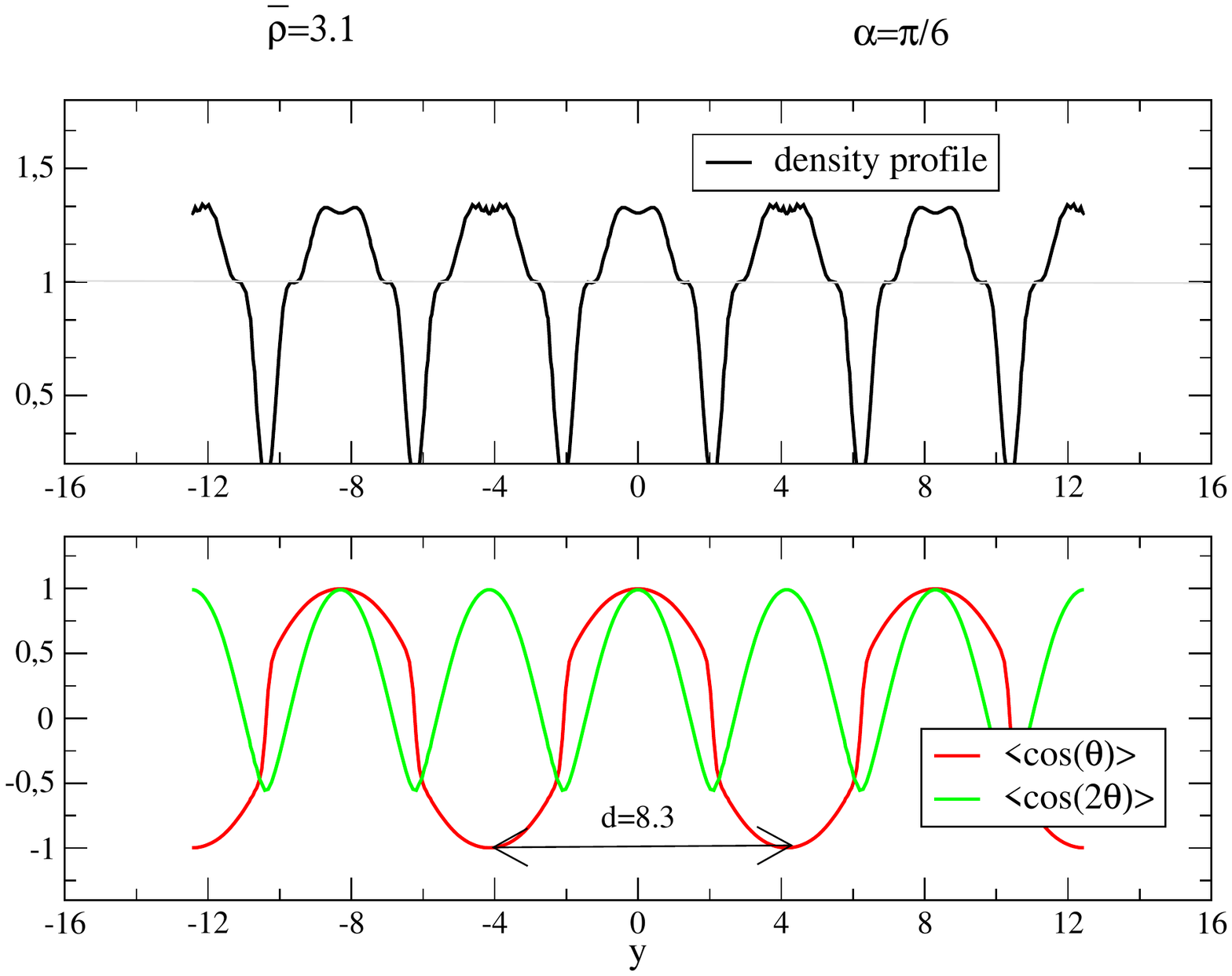}
\vspace{0.4cm}
\includegraphics[angle=90,width=0.666\columnwidth]{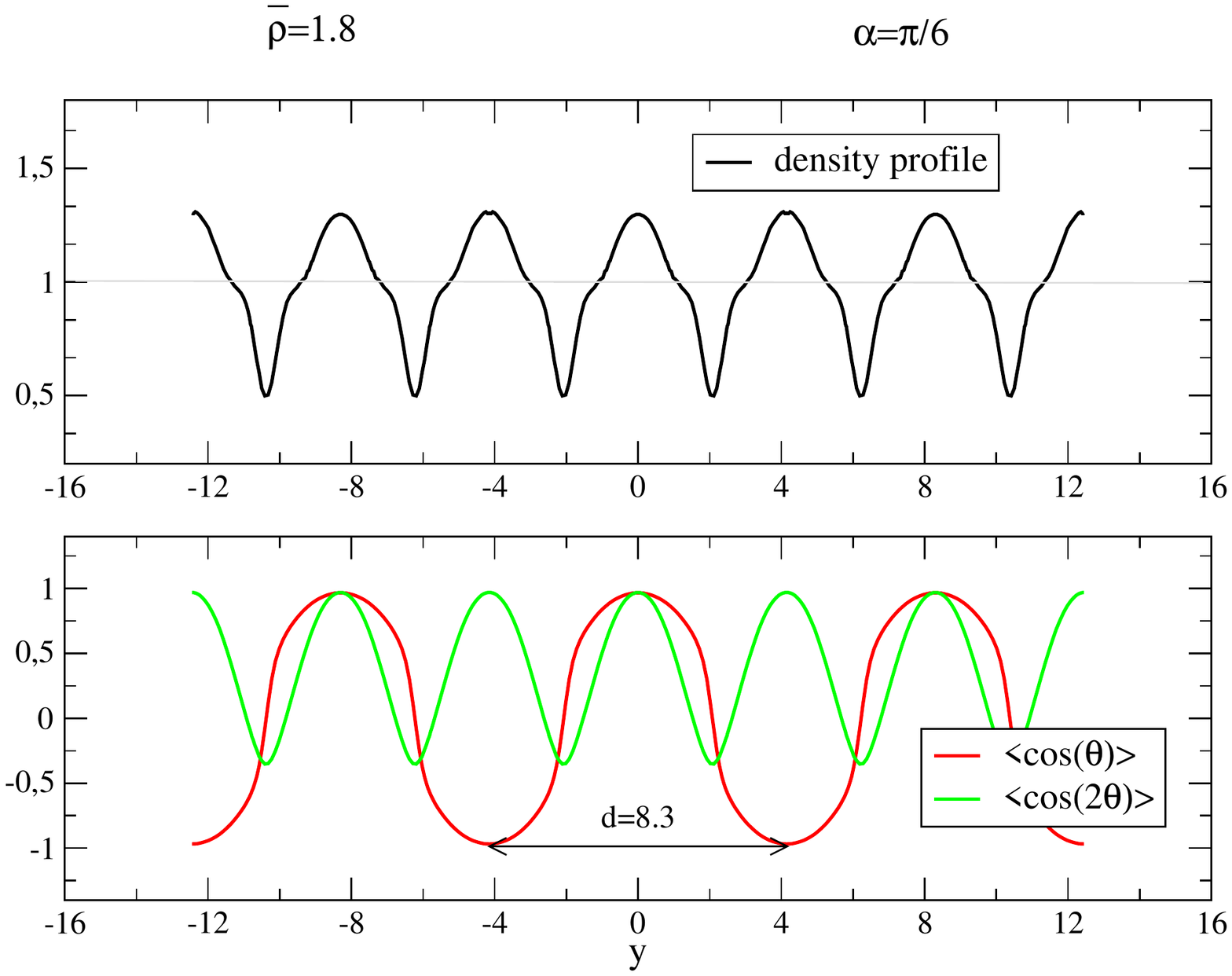}
\end{center}
\caption{Density profile $P(y)$  and order parameters 
(calculated with respect to the $y-axis$ of laboratory frame) for the 
$S_{SB}$ phase 
obtained for  ${{\overline{\rho}}=3.1}$ for \emph{cis} molecules characterized 
by the angle $\alpha=\pi/6$ (Fig.2). 
Nematic order parameter 
$2\langle (\mathbf{b}\cdot \mathbf{\hat y})^2 \rangle -1$ 
measured in the LAB
is given by green colour and polar order parameter 
$\langle \mathbf{a}\cdot \mathbf{\hat x} \rangle$  by red colour. 
Alternating positive and negative parts of the polar order and the pitch of polar modulation 
is twice the period of density modulation, and of about 3 times molecule's length 
is indicative of the splay bend formation. } 
\label{ro31}
\end{figure}

We start an illustrative characteristic of $S_{SB}$ 
with Fig.~\ref{ro31} showing the exemplary results for the density profile and for the averages
$\left<cos2\theta\right>=
2\langle (\mathbf{b}\cdot \mathbf{\hat y})^2 \rangle -1$ and 
$\left<cos\theta\right>=\langle \mathbf{a}\cdot \mathbf{\hat x} \rangle$
obtained for two reduced densities
of $3.1$ and $1.8$ and for $\alpha=\pi/6$ (\emph{see} Fig. 2).
The unit vectors $\mathbf{\hat x}$ and $\mathbf{\hat y}$ are, respectively,  
parallel and perpendicular
to the $y$ axis of the laboratory frame (LAB), while
the layers are perpendicular to $y$ (parallel to $x$).
 These averages show the modulation 
 of the density along with periodicity 
 in the 
 orientational degrees of freedom. The first of the two averages
corresponds to the component of 
the nematic order parameter measured in the LAB, while the order parameter 
$\langle \mathbf{a}\cdot \mathbf{\hat x} \rangle$ 
is the average molecular polarization measured with respect to the x-axis of the LAB.
The structure thus obtained is modulated and consists of layers, which makes it recognized 
as a smectic with a layer spacing (as seen in the density modulation) of approximately $d=4.15$. 
The pitch (approximately three molecular lengths) of the orientational modulation 
is twice the length of the density modulation, which, 
taking into account the behavior of the polar order parameter (red curve),  
is typical for splay bend deformations. 
The values of the nematic order parameter ($\left<cos2\theta\right>$) approach 
unity within the density layers 
and reach the value of $-0.5$ 
at the edges of the density layer.
The profile of the polar order parameter ($\left<cos\theta\right>$) consists of 
identically shaped parts that alternate between positive and negative values.
Within the layers, the density profile contains parts where the density 
is at a constant level, 
but the range of them is small. 
Slight
fluctuations in the density curve are a numerical artifact. This we know since if a 
smaller number of Gaussian points is used, then they are getting much larger.  
At the same time, the orientational order parameters are not so sensitive 
to the accuracy of numerical integration.

\begin{figure}[htb]
\begin{center}
\vspace{0.4cm}
\includegraphics[angle=90,width=0.8\columnwidth]{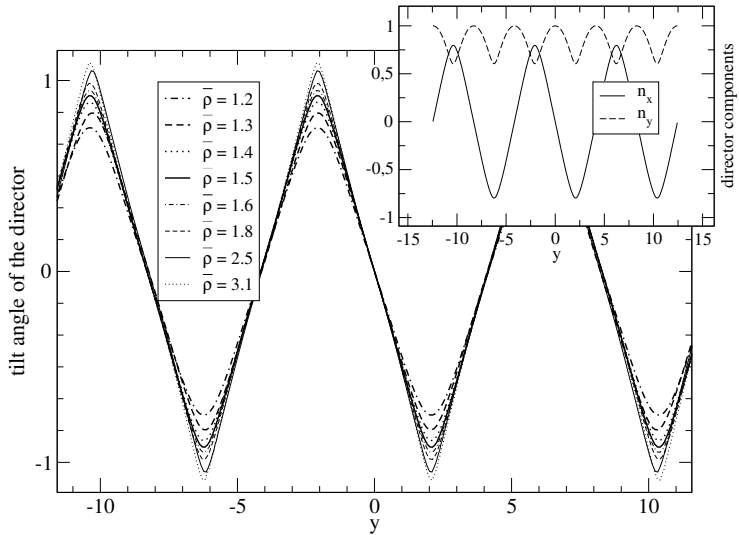}
\end{center}
\caption{Reduced density dependence of tilt angle of the director 
with respect to the layer normal 
(parallel to the y-axis of laboratory system of frame) 
as calculated from the DFT theory for  the  
$S_{SB}$ phase. Inset shows exemplary components of the director for
$\overline{\rho}=1.5$.} 
\label{AlfaAng}
\end{figure}

\begin{figure}[htb]
\begin{center}
\vspace{0.4cm}
\includegraphics[angle=90,width=0.8\columnwidth]{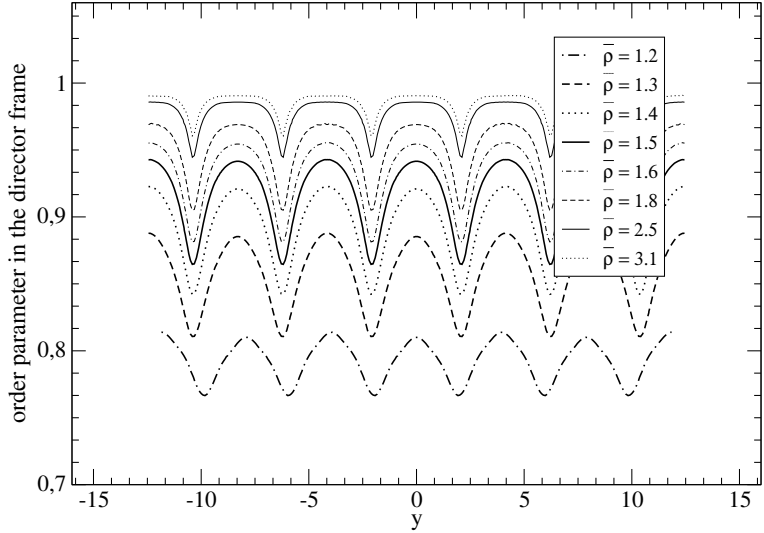}
\end{center}
\caption{Nematic order parameter profiles $2\langle (
\mathbf{\hat n}(y)\cdot
\mathbf{b})^2\rangle -1$   in the director frame for the 
$S_{SB}$ phase for different densities.  } 
\label{Qrho}
\end{figure}

\begin{figure}[htb]
\begin{center}
\vspace{0.4cm}
\includegraphics[angle=90,width=0.8\columnwidth]{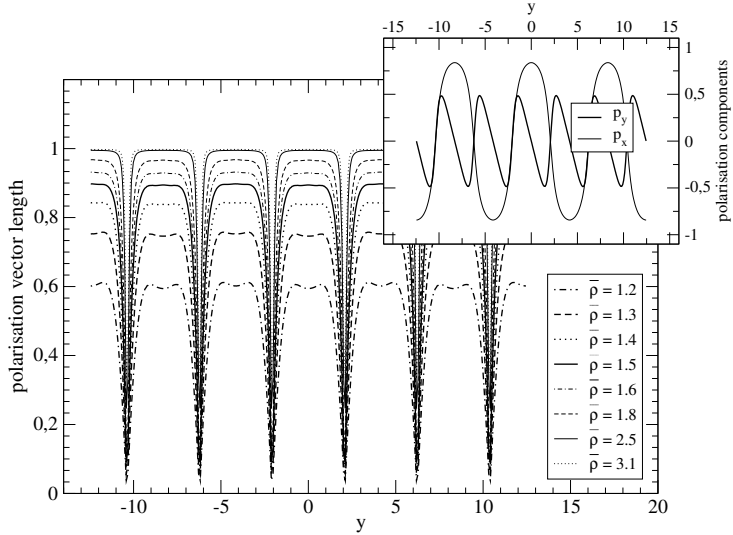}
\end{center}
\caption{
Length of the polarisation vector $|\langle 
\mathbf{\hat m}(y)\cdot
\mathbf{a}\rangle|$ for different densities. 
Inset shows exemplary components of the polarization vector for $\overline{\rho}=1.4$. 
Please remind that average polarization is always  perpendicular to the director for $S_{SB}$. 
} 
\label{polarvector}
\end{figure}

In Fig.\ref{AlfaAng}, we present the density dependence 
of the director tilt angle 
(defined as the average orientation 
of the central molecular segment) for the phase $S_{SB}$, further supporting the hypothesis that 
the smectic phase is of the splay-bend type. 
Within the smectic layer it exhibits a practically linear dependence with distance,
which is in agreement with the Monte Carlo simulations
in Fig.~\ref{fig:order_profile}.
This result is of special importance because it is also 
the main assumption used to estimate the nematic-smectic splay bend 
boundaries for the full spectrum of conformers in the next subsection.
In Fig.~\ref{Qrho} we present the corresponding nematic order parameter profiles 
$<2(\mathbf{\hat n}(y)\cdot \mathbf{b})^2-1>$ 
calculated in the director frame for different densities. 
These profiles have been obtained from 
the 2D local alignment tensor $\mathbf{Q}(y)$, whose components in the laboratory frame
are given by
\begin{equation} \label{Qtens}
 Q_{ij}=<2 b_i b_j-\delta_{ij}>=n_in_j<2(\mathbf{\hat n}\cdot \mathbf{b})^2-1> +
m_i m_j<2(\mathbf{\hat m}\cdot \mathbf{b})^2-1>. 
\end{equation}
Here $\mathbf{\hat n}$ and $\mathbf{\hat m}$ of the components $n_i$, $m_i$, respectively,
are local orthonormal eigenvectors of $\mathbf{Q}$ where the positive eigenvalue is associated with
the director $\mathbf{\hat n}$. Unlike the $\mathbf{\hat n}\equiv -\mathbf{\hat n}$ symmetry
of the director, $\mathbf{\hat m}$ is the polar vector, reflecting the polar nature of $S_{SB}$. 
As can be seen in Fig.~\ref{Qrho} not only are the eigenvectors of 
$\mathbf{Q}$ positionally dependent, but the nematic order parameter profile 
$<2(\mathbf{\hat n}\cdot \mathbf{b})^2-1>$ 
exhibits weak modulation that decreases the more ordered phase for larger densities. 
However, the local orientational order of $S_{SB}$ is quite high.
In Fig.~\ref{polarvector}  the local length $0 \le| <\mathbf{a}\cdot \mathbf{\hat m}>| \le 1$ 
of the average polarization vector (which stays parallel to $\mathbf{\hat m}$)
is presented for various reduced densities. Again, we observe a high, slightly modulated 
behavior of this observable in $S_{SB}$ of periodicity consistent with that of the nematic order parameter. 
However, the periodicity of the average polarization vector, alternating with discontinuity
between the positive and negative x-component, 
is twice that of the director (\emph{see} inset in Fig.(\ref{polarvector})). 
Furthermore, the transition temperature $N-S_{SB}$ for pure \emph{cis} molecules 
is lower than in the simulations. A major cause of that is the neglect of 
the \emph{zigzag} conformers. 
The value of the pitch $d=8.3$ is nearly constant for a wide 
range of densities. Only for the cases close to the $N-S_{SB}$ transition, 
which takes place at a density of about $\bar{\rho}=1.05$, it becomes slightly lower.
For example, for $\bar{\rho} = 1.2$ the pitch value is $d = 7.9$.

\begin{figure}[t]
\begin{center}
\vspace{0.8cm}
\includegraphics[angle=90,width=0.8\columnwidth]{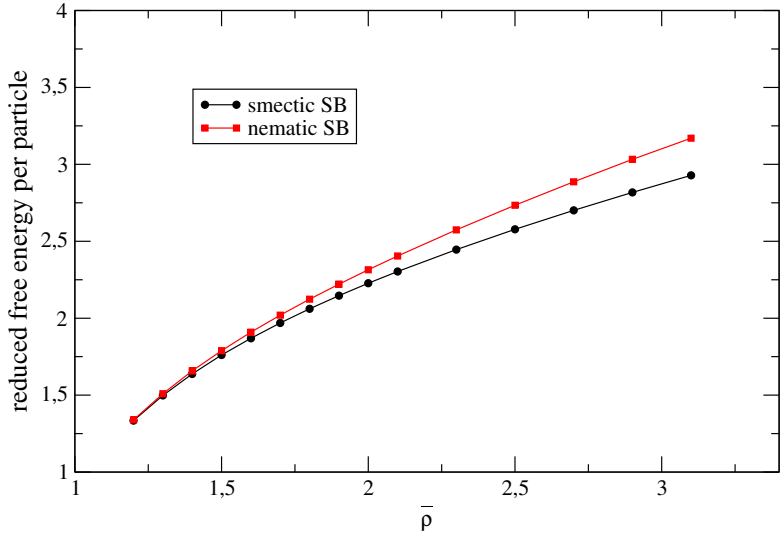}
\end{center}
\caption{Reduced free energy per particle versus average density $\overline{\rho}$ 
for smectic- and nematic splay bend phases as obtained from DFT for \emph{cis} molecules 
with arms titled at 
angle $\pi/6$.
} 
\label{energie}
\end{figure}

The DFT splay-bend solutions for the system studied always exhibit a spatial modulation 
and are thus recognized as $S_{SB}$. 
However, with the ansatz $P(y)=const $ imposed on (\ref{freeener}) it is still possible to obtain 
a stationary solution of Eq.~(\ref{freeener}) which is the pure 
nematic splay bend.
Effectively, this can be achieved by using the following ansatz
in the iteration process:
\begin{equation}
P(y,\theta)=\frac{P(y,\theta)}{\int P(y,\theta) d \theta},
\label{ansatz_nsb}
\end{equation}
along with the minimization of (\ref{Heff13}) with respect to period $d$.
As a result, one can directly compare the free energies per particle 
of $N_{SB}$ and $S_{SB}$.
This is shown in Fig.~\ref{energie} where the free energy of 
$S_{SB}$ is found to be systematically lower than that of $N_{SB}$. 
With a decrease in the reduced density, the departure of the free energy of $S_{SB}$ 
from that of $N_{SB}$ becomes smaller and almost disappears near 
the $N-S_{SB}$ phase transition. However, no transition $N_{SB}-S_{SB}$ has been detected.
\subsection{Bifurcations from isotropic and nematic phases}
Our last step will be to use the full DFT formalism (\ref{freeener}) for our model particles 
(Fig. \ref{fig1})
and to find the ordered phases that can bifurcate from the isotropic phase in the $(\alpha, \rho)$
plane. We will also estimate the phase boundaries in this plane between 
the nematic and smectic splay-bend assuming that the nematic phase is perfectly aligned.

The bifurcation scheme applied to equations such as (\ref{self1}) has been 
discussed in detail in a series of publications \cite{bif1_Kayser,bif2_Mulder,longa_bif1,longa_bif3,longa_bif4,longa_bif2}.
Here, we generalize these results to the case with conformational degrees 
of freedom. We assume that the reference state is characterized by the distribution
function $P_0(X)$. Then, at the bifurcation from state $P_0(X)$ to a new state $P(X)$
the difference $P(X)-P_0(X)$ is arbitrarily small,
allowing one to linearize Eq.(\ref{self1}) about $P_0$.
Simulations show that in the isotropic phase, all conformers are equally populated,
allowing us to take $P_0(X)=P_0(s)\equiv P_{I}=const$. Similarly, we assume that 
$P_0(X)$ is known for ideally oriented molecules in the nematic phase.
If we are only interested in finding the densities of bifurcations, it is sufficient to analyze 
the linearized form of Eq.~(\ref{self1}) given by
\begin{equation}
 P(X) = Z_0^{-1}\left(1-\overline{\rho}S\underset{(X_2)}{{Tr}}\left\{
\Theta \left[ \xi (X,X_{2})-r_{12}\right]\, P(X_{2}) \right\}\right),  \label{linearization}
\end{equation}
with $Z_0$ calculated for $P_0$.
For the analysis of liquid crystalline phases (with the exception of $S_{SB}$), we decompose the probability distribution function $P(X)$ into a complete and orthonormal 
basis $F_{\alpha}$
\begin{equation}
P(y,\theta,\mathbf{s})=\sum_{\alpha}^{ }\langle F_{\alpha} \rangle F_{\alpha}(y,\theta,\mathbf{s}),
\label{one_particle_expansion}
\end{equation}
where $F_{\alpha}(y,\theta,\mathbf{s})$ is the product of orthonormal 
functions in separate subspaces. It reads:
\begin{equation}
F_{\alpha}(y,\theta,\mathbf{s})\equiv
F_{abkl\sigma}(y,\theta,\mathbf{s})=n_{abkl\sigma}\psi_a(k y)\phi_b(l\theta)f_{\sigma}(\mathbf{s}),
\label{rownanie04}
\end{equation}
where $n_{abkl\sigma}$ is the normalization constant and where discrete indices $\{a,b,k,l,\sigma\}$
are sequentially numbered with a single index $\alpha$, where low $\{k,l,\sigma\}$s correspond to low values of $\alpha$.
The spatial y-dependent functions $\psi_a(ky)$ are discrete Fourier cos and sin modes    
\begin{equation}
\psi_a(ky)=\left\{\begin{matrix}
\cos(\frac{2 \pi k y}{d}), & a=1\\ 
\sin(\frac{2 \pi k y}{d}), & a=2,
\end{matrix}\right. 
\label{rownanie05}
\end{equation}
where  $k=0, 1, 2,...$,  and where $d$ is the periodicity of the structure
($S=L^2,\, L=Md,\, M \in \mathbb{N}$).
Angle-dependent functions $\phi_b(l\theta)$ are
\begin{equation}
\phi_b(l\theta)=\left\{\begin{matrix}
\cos(l\theta), & b=1 \\ 
\sin(l\theta), & b=2 ,
\end{matrix}\right.
\label{rownanie06}
\end{equation}
where $l=0, 1, 2,...$.
Finally, the four purely conformational functions 
$f_{\sigma}(\mathbf{s})\equiv f_{\sigma}(s_1,s_2)$ 
 are given by
\begin{equation}
f_{\sigma}(s_1,s_2)=\left\{\begin{matrix}
\frac{1}{2}, & \sigma=0 \\ 
\frac{s_1+s_2}{2\sqrt{2}}, & \sigma=1 \\ 
\frac{s_1-s_2}{2\sqrt{2}}, & \sigma=2 \\ 
\frac{1}{2}s_1s_2, & \sigma=3\, .
\end{matrix}\right. 
\label{rownanie0101}
\end{equation}
The normalization constant $n_{abkl\sigma}$ is such that
\begin{equation}
Tr_{(X)}\left[F_{abkl\sigma}\left(y,\theta,\mathbf{s} \right )F_{a'b'k'l'\sigma'}\left(y,\theta,\mathbf{s} 
\right ) \right ]=\delta_{aa'}\delta_{bb'}\delta_{kk'}\delta_{ll'}\delta_{\sigma\sigma'},
\label{rownanie10}
\end{equation}
which allows identification: $\langle F_\alpha \rangle= Tr_{(X)}\left[P(y,\theta,\mathbf{s})F_{\alpha}\left(y,\theta,\mathbf{s} 
\right ) \right ]$. Substituting (\ref{rownanie04}) into
(\ref{rownanie10}) explicitly gives
\begin{equation}
n^2_{abkl\sigma}=\frac{2}{\pi L}\left[1-\left(\frac{1}{2}\right)^{2-a} \delta_{k0} \right ] 
\left[1-\left(\frac{1}{2}\right)^{2-b} \delta_{l0} \right ].
\label{rownanie14}
\end{equation}

The first term $\langle F_{\alpha=0} \rangle = F_{\alpha=0}(y,\theta,\mathbf{s}) 
\equiv \langle F_{11000}\rangle = F_{11000}(y,\theta,\mathbf{s})$ in the expansion
(\ref{one_particle_expansion}) is the normalization constant of $P(X)$. The remaining 
terms $\langle F_{\alpha \ne 0}\rangle$ are the order parameters. For example,
$\langle F_{1102\sigma}\rangle$ are the leading order parameters of nematics.
Nonzero $k$s indicate in the smectic ordering, which can be non-polar 
(only even $l$s are present)
or polar (at least $l=1$ is present). 

The index $\sigma$ classifies the order parameters 
according to the local($k\ne0$)/global($k=0$) population of various conformational
states. More specifically,
for a structure of $\sigma=0$, each conformer population is represented with equal probability. 
The order parameters with $\sigma=1$ measure the polar order, and they vanish for 
chiral conformers. In contrast to case $\sigma=1$, the order parameters with $\sigma=2$ are 
responsible
for the chiral order, since there are only chiral conformers that contribute to this.
Finally, the order parameters with $\sigma=3$ measure the relative local/global importance 
of chiral and non-chiral conformers for a given phase. If we are interested 
in the (local) concentration  
of a given fraction of conformers, these are found by performing a simple average
 over $P(X)$.
For example, the concentration of the fraction \emph{zigzag+} is given by
\begin{equation}
  \langle \delta_{s_1,1} \delta_{s_2,-1} \rangle = \frac{F_{abkl0}}{4} + \frac{\langle F_{abkl2}\rangle}{\sqrt{2}}
  -\frac{\langle F_{abkl3} \rangle}{2},
\end{equation}\label{zigzagfraction}
where $\delta_{i,j}$ is the Kronecker delta.

Using the definitions introduced above, we can now proceed with the identification
of states that bifurcate from the reference one. In the first step, we expand the 
kernel $Z_0^{-1} S \Theta \left[ \xi (X^{'}_{1},X^{'}_{2})-r^{'}_{12} \right]$ in the basis
$F_{\alpha}(X)$, substitute the expansion for the equation (\ref{linearization}), multiply both 
sides by $F_{\gamma}(X)$ and trace over $X$.
This procedure reduces Eq.~(\ref{linearization}) to a set 
of linear equations for $\langle F_{\gamma} \rangle$
\begin{equation}
   \langle F_{\gamma} \rangle  = \langle F_0 \rangle \delta_{\gamma 0} - \overline{\rho}\sum_{\beta} \Theta_{\gamma\beta} 
    \langle F_{\beta} \rangle, \label{P_with_theta_decomposed}
\end{equation}
where
\begin{equation}
 \Theta_{\alpha\beta}=    Z_0^{-1} S\underset{(X^{'}_{1}X^{'}_{2})}{{Tr}}\left\{ F_{\alpha}(X^{'}_{1}) \Theta \left[ \xi (X^{'}_{1},X^{'}_{2})-r^{'}_{12} \right]F_{\beta}(X^{'}_{2}) \right\},
\end{equation}
and where the above calculations are limited to
the $n \times n $ matrix $\Theta_{\alpha\beta}$,
corresponding to the leading terms in (\ref{rownanie04}), usually 
of $k\le 2$ and $l \le 2$, which generally makes the 
linear equation (\ref{P_with_theta_decomposed})
of order $n\le 100$. 

The next step is diagonalization of the matrix $\Theta_{\alpha\beta}$. Let 
$\Theta_{\alpha\beta}$=$ p_{\alpha \mu} 
\lambda^{\theta}_{\mu} (\mathbf{p}^{-1})_{\mu\beta}$ where $\lambda^{\theta}_{\mu}$
are eigenvalues of 
$\Theta_{\alpha\beta}$ and $\mathbf{p}$ transforms $\Theta_{\alpha\beta}$ 
into diagonal form. With the aid of $\mathbf{p}$, equation (\ref{P_with_theta_decomposed})
can be rewritten in a simpler form. It reads: 
\begin{equation}
\langle\chi_{\alpha}\rangle= (\mathbf{p}^{-1})_{\alpha 0}\langle F_0 \rangle
-\overline{\rho}\lambda^{\theta}_{\alpha} \langle\chi_{\alpha}\rangle,
\label{diagonalForm}    
\end{equation}
where $\chi_{\alpha}=\sum_\beta (\mathbf{p}^{-1})_{\alpha \beta} F_{\beta}$.
Now, candidates $\{\chi_{\mu}\}$ for the bifurcating states are those that do not couple with
$\langle F_0 \rangle$ \emph{i.e} for which $(\mathbf{p}^{-1})_{\mu 0}\langle F_0 \rangle=0$.
The second condition is that the corresponding $\{\lambda^{\theta}_{\mu}\}$ are real negative numbers.
With these restrictions, we are left with a set of homogeneous equations 
$\,\{\langle\chi_{\mu}\rangle=
-\overline{\rho}\lambda^{\theta}_{\mu} \langle\chi_{\mu}\rangle\,\}$ for 
$\langle\chi_{\mu}\rangle\ $,
which produce non-zero solutions for $\langle\chi_{\mu}\rangle$ given 
that $\overline{\rho}\ge \overline{\rho}^{*}_{\mu} $, where 
$\overline{\rho}^{*}_{\mu}=-\frac{1}{\lambda^{\theta}_{\mu}}$.
The first bifurcating state $\overline{\rho}^{*}_{\mu_0}$ corresponds 
to the minimal 
density among $\{\overline{\rho}^{*}_{\mu}\}$: $\overline{\rho}^{*}_{\mu_0}=\mathrm{Min}\{\overline{\rho}^{*}_{\mu}\}$.
The corresponding bifurcating probability density distribution function is given by 
$P(X)= P_0(X) + \epsilon \chi_{\mu_0}(X)$, where arbitrarily small $\epsilon\equiv \langle\chi_{\mu_0}\rangle $, can be determined with the higher order 
perturbation calculations.
\subsubsection{Bifurcations from isotropic phase}
\begin{figure}[htb]
\begin{center}
\includegraphics[width=0.75\columnwidth]{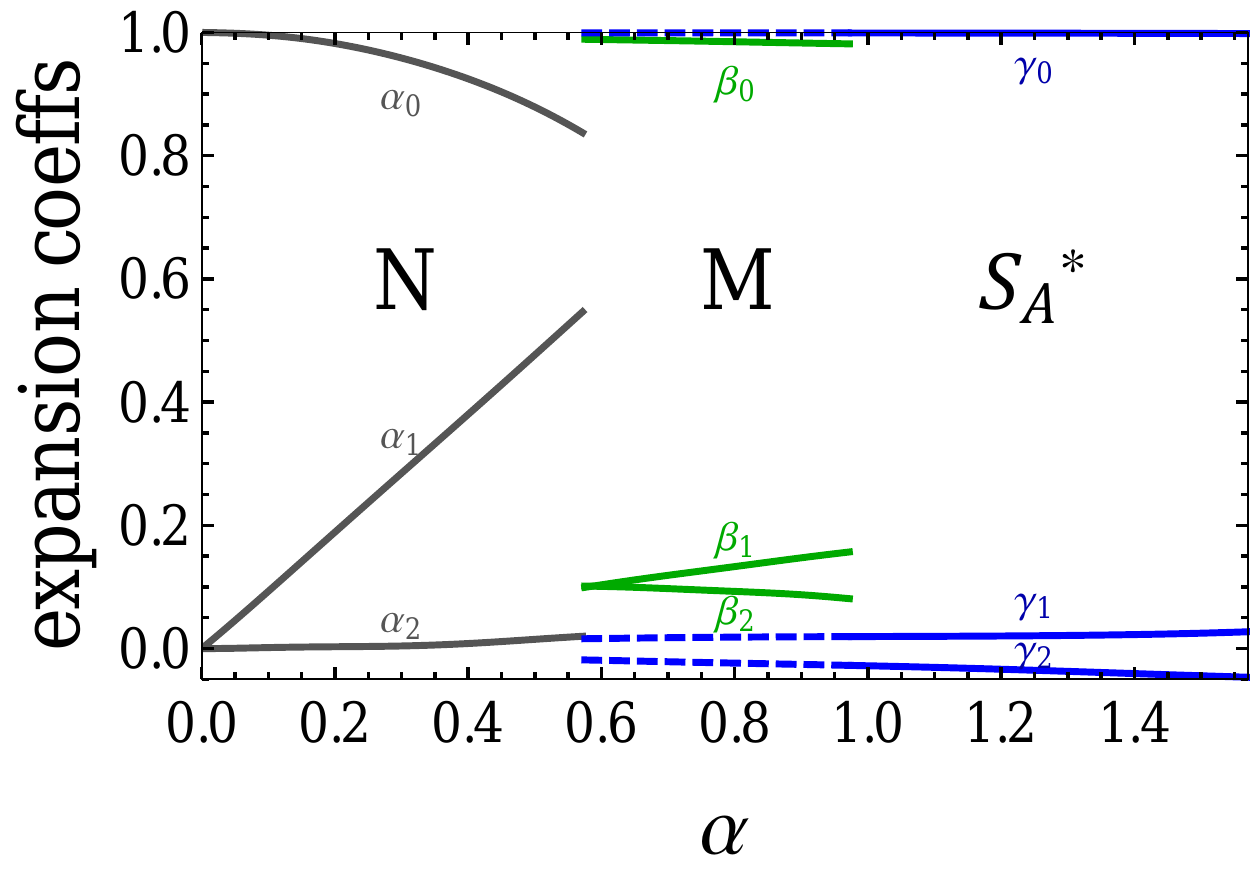}
\end{center}
\caption{Coefficients of decomposition of  bifurcating states in the $F_\gamma$-basis
for $N$, $M$ and $S_A^*$ structures.}
\label{fig:order_parameters}
\end{figure}

We found the bifurcation densities from the isotropic phase to the different nematic 
and smectic phases using a full set of degrees of freedom. 
In our analysis $k=1$, $a=1$ or $2$ (Eq. \ref{rownanie05}), 
$l$ ranges from $0$ to $2$ for $b$ equal to $1$ or $2$ (Eq. \ref{rownanie06}), 
and $\sigma$ varies from $0$ to $3$ (Eq. \ref{rownanie0101}). 
This limits the size of the matrix $\Theta_{\alpha\beta}$ for the bifurcation analysis
to $40$ by $40$. Due to the dependence of $\Theta_{\alpha\beta}$ on the wave vector, 
for any specific opening angle $\alpha$ of particles, Fig.~(\ref{fig1}), we 
diagonalize a series of matrices, each for a different wave vector, and find the eigenvalue
that gives the lowest bifurcating density. For diagonalization, we used 
the Jacobi method involving Givens rotations, since these matrices were all 
symmetrical. The results are presented in Fig. \ref{fig:phase_diagram}, where the 
dark gray line corresponds to the bifurcation from the isotropic phase 
($I$) to (chiral) nematic ($N$) phase, green line from the isotropic phase 
to the so-called mixed phase (M) and the blue line from isotropic 
to chiral smectic A ($S_A^*$). 
The corresponding bifurcation probability densities $P(y,\theta,\boldsymbol{s})$ are:

\emph{(a)} for the $I-N$ bifurcation (dark gray line):
\begin{equation}
  \begin{split}
    P(y,\theta,\boldsymbol{s}) \approx & 
    \frac{1}{8\pi S}+ \varepsilon [
    \alpha_0 \cos (2 \theta ) f_0(s_1,s_2) \\
    &
    + \alpha_1 \sin (2 \theta ) f_2(s_1,s_2)
    + \alpha_2 \cos (2 \theta ) f_3(s_1,s_2) + ...\,
    ] .
  \end{split}
\end{equation}
The coefficients $\{ \alpha_i,i=0,1,2\}$ are presented in Fig.~\ref{fig:order_parameters} 
in dark gray, similar to that of the phase diagram in Fig.~\ref{fig:phase_diagram}.
The leading term $\alpha_0 \cos (2 \theta ) f_0(s_1,s_2)$ represents the 
nematic order with an equal fraction of different conformers. The next  
term $\alpha_1 \sin (2 \theta ) f_2(s_1,s_2)$ becomes relevant for $\alpha > 0$. 
It monitors the appearance and increase of chiral fractions (zigzags) with increasing $\alpha$,
with a maximum fraction achieved when the central molecular segment is oriented
at $\theta = \pm \pi/4$ with respect to the director. This means that the overall 
nematic phase becomes macroscopically chiral. At the same time, the concentration 
of \textit{cis}-type particles along the director outweighs the concentration of 
the chiral conformers ($\alpha_2 > 0$), in agreement with Monte Carlo simulations,
Fig. \ref{fig:eosandconfs};

\emph{(b)}
for the $I-M$ bifurcation (green line: $0.18 \pi \lesssim \alpha \lesssim 0.31 \pi$), where 
$M$ stands for the 'mixed phase', the situation is more complex because 
the two different smectic phases of nearly equal bifurcation densities 
compete (green line, dashed blue line). The phase of slightly lower density 
is given by the probability density:
\begin{equation}
    \begin{split}
    P(y,\theta,\boldsymbol{s}) \approx & \frac{1}{8\pi S} +\varepsilon  [
    \beta_0 \sin\left(\frac{2\pi y}{d}\right) f_0(s_1,s_2)
    +\beta_1 \cos\left(\frac{2\pi y}{d}\right) \sin(\theta)  f_1(s_1,s_2) \\
    & 
    +\beta_2 \sin\left(\frac{2\pi y}{d}\right) \cos(2\theta) f_0(s_1,s_2) + ...]. 
    \end{split} \label{second_bifurcation}
\end{equation}
The structure is of the smectic type ($\beta_0>0$). It has a period of approximately 
$d\approx 10.1$ 
for all opening angles $\alpha$ of the particles and is characterized by 
the coefficients $\beta$ as shown in Fig. \ref{fig:order_parameters}. Given 
the period of the structure (roughly three molecular lengths) and the polarization 
wave of the \emph{cis}-conformers  ($\beta_1 \ne 0$) it can be identified 
as a non-chiral smectic splay-bend ordering;

\emph{(c)} for the $I-S_A^*$ bifurcation (blue line):
\begin{equation}
    \begin{split}
        P(y,\theta,\boldsymbol{s}) \approx & \frac{1}{8\pi S} +\varepsilon [
    \gamma_0 \cos \left(\frac{2\pi y}{d}\right)f_0(s_1,s_2) +\gamma_1 \cos \left(\frac{2\pi y}{d}\right) \sin (2 \theta )f_2(s_1,s_2)  \\ 
    & +\gamma_2  \sin \left(\frac{2\pi y}{d}\right) \sin (\theta )f_1(s_1,s_2) + ...].
    \end{split}\label{third_bifurcation}
\end{equation}
The period of this structure is practically constant and equal to $d=4.25$ for 
all opening angles of the particles.
The coefficients $\{ \gamma_i,i=0,1,2\}$ are shown in Fig. \ref{fig:order_parameters} 
as continuous blue lines. Here, the dominant
$\gamma_0$-term  is of a smectic A type. Furthermore, the presence of the chiral term, 
proportional to $\gamma_1$, makes \textit{zigzag} molecules preferred in this phase.
These results remain in agreement with the simulations and allow us to identify this 
phase as chiral smectic A ($S_A^*$). 
They also explain why it is so difficult to unequivocally interpret the $M$ structure in
Monte Carlo simulations. The reason is that the dashed blue 
line representing bifurcation to $S_A^*$
in the phase diagram, Fig.~\ref{fig:phase_diagram}, nearly overlaps the green line that describes
the competing smectic structure (\ref{second_bifurcation}).
\subsubsection{Bifurcations from ideally ordered nematic phase}
Monte Carlo simulations and the results of previous sections revealed that at the 
transition from nematic to smectic splay bend the particles are 
highly ordered about the director. Therefore, to study a bifurcation from an ordinary nematic phase
to more ordered phases, we can assume the perfect nematic 
order approximation \cite{karbowniczek}, where the central molecular 
segments always stay parallel to the director 
($\mathbf{\hat n}(y)\cdot \mathbf{b_i}=1,\, i=1...N $).
Now, we can proceed as before by expanding $P(y,s)$ in the basis 
$F_{abk0\sigma}(y,\theta,\mathbf{s})$, where in the ideal nematic order 
approximation the terms $\phi_b(l\theta)$ are dropped. 
As a consequence, the problem is reduced to the diagonalization of 20 by 20 matrices 
for the period $d$, which minimizes the density of the bifurcation. 
Results are shown as a red line in the phase diagram.
\section{Conclusions}
Since the seminal work of Onsager on steric interactions in nematics \cite{onsageraartykul}, 
simulations and theoretical works that invoke the packing entropy provided important 
insights into the self-assembly of densely packed systems, especially liquid crystals. 
Without escaping from the main idea that molecular shape is the primary
factor in the formation of different liquid crystalline structures,
our attempt here was to add a new factor, namely conformational degrees of freedom. 
We used simple three-segment calamitic particles under 
strong planar confinement to probe the structural behavior of the LC samples. 
We assumed four predominant conformations per molecule 
\textit{cis$\pm$} and \textit{trans$\pm$} ones. Therefore, the model
served as the simplest example in which an interplay between packing entropy and 
sterically restricted conformational entropy can be studied in detail.
So far only separate systems, composed of 
\textit{cis$+$} or \textit{trans$+$} rigid trimers were considered.  

We were particularly interested in the formation of splay-bend and chiral ordering 
previously reported for \textit{cis$+$}  and \textit{trans$+$}, respectively.
Using constant pressure Monte Carlo simulations and Onsager-type density functional theory, 
we showed that the presence of conformational degrees of freedom leads to a rich spectrum
of stable phases where different average fractions of conformers combine to determine properties
of the resulting phases. While in the isotropic phase all fractions are equally populated
and the phase is nonchiral, the ordered phases that are formed from the isotropic phase 
are all macroscopically chiral.
The chirality of ordered phases is a function of density and opening angle,
and the study of the average fraction of the \textit{trans-} and \textit{cis-} 
conformers revealed that the nematic and splay-bend phases 
are dominated by the achiral \emph{cis} conformers, while the chiral \emph{zigzag}s 
prevail in the $S_A^*$ phase. 
Hence we observe conformationally and sterically induced isotropic-nematic- and isotropic-smectic
chiral symmetry breaking (CSB). In both, the nematic and 
smectic $A^*$ phases domains of opposite chiralities are formed as a manifestation of CSB.

We obtain a bifurcation from the isotropic to the nematic phase 
and two bifurcations from the isotropic to the smectic phases without an intermediate 
nematic phase. We also find that in both smectic phases, the period $d$ of the smectic structures  
at bifurcation does not depend on the opening angle and is equal to $10.1$  in $M$ 
and $4.25$ in the $S_A^*$ phases. Surprisingly good  
qualitative agreement of these results 
with the constant-pressure Monte Carlo simulations
is obtained along with structure identification. Furthermore, difficulties in stabilizing 
the $M$ phase in MC simulations are explained by the proximity 
of a competing $S_A^*$ phase, as seen in the bifurcation analysis of the corresponding 
Onsager model. 

  The most interesting observation is an identification of the smectic splay-bend phase. 
 Although $S_{SB}$ is dominated by the $cis\pm$ conformers, the $zigzag\pm$ ones 
 are also present, occupying with maximal probability the places where 
 the steric polarization wave changes sign.
 They seem to help in stabilizing the smectic ordering.
 As stated in the Introduction, the concurrent, long-sought splay-bend nematic phase 
 has been detected recently in colloidal systems of bent-core particles. 
 However, a novel grand potential Landau-deGennes theory as well as simulations 
 suggest that the alleged splay-bend phase stabilizing in colloidal systems to 
 display density modulations. Our flexible trimers are also forming stable $S_{SB}$.
 Therefore, it seems that the stable splay-bend phase appears to have the 
 key symmetry of smectics. However, the bifurcation analysis does not exclude
 the pure nematic splay-bend solutions.

The structures obtained in the simulations do not have a true long-range order; 
however, the nematic of the particles with a small angle $\alpha$ and smectics 
seems to form domains with the largest correlation length. 
We showed that despite the plethora of functions used in the literature, $g_2(r)$ 
is a sufficiently good indicator of (quasi-) liquid crystal phase transitions 
for the particles studied. We also provided a detailed theoretical description 
of the results obtained in the simulations.

  Finally, the subject of molecules with conformational degrees of freedom has 
been addressed by Tavarone \emph{et. al.}\cite{tavarone2} in their 
kinetic Monte Carlo modeling of conformational changes taking place 
when photo-switchable molecules of the derivative of the dye Methyl-Red 
are attached to a surface and subjected to linearly polarized light.
In the cited work, the generated photo-induced change in molecular shape was 
represented by straight and bent needles.

   Our model, which takes into account equilibrium \textit{trans----cis}, but ignores
the monomer fraction and assumes that open and cyclic acid trimers are formed 
instantaneously and with equal probability, can also be viewed as a simplified 
description of the hydrogen bond dissociation-association process. We think that at least
some of the steady states \textit{trans----cis} of the complexes should be qualitatively 
similar to the equilibrium configurations predicted by this work. 
As such studies have not been explored in monolayer liquid crystal science, it would be interesting 
to see how more complex scenarios, \emph{e.g.} with monomers and energetic factors describing 
complex formation influence the resulting entropy-induced orientational order. 

\section*{Acknowledgments}

This work was supported by Grant No. DEC-2021/43/B/ST3/03135 of the National Science Centre in Poland.

\bibliography{main}

\begin{thebibliography}{60}%
\makeatletter
\providecommand \@ifxundefined [1]{%
 \@ifx{#1\undefined}
}%
\providecommand \@ifnum [1]{%
 \ifnum #1\expandafter \@firstoftwo
 \else \expandafter \@secondoftwo
 \fi
}%
\providecommand \@ifx [1]{%
 \ifx #1\expandafter \@firstoftwo
 \else \expandafter \@secondoftwo
 \fi
}%
\providecommand \natexlab [1]{#1}%
\providecommand \enquote  [1]{``#1''}%
\providecommand \bibnamefont  [1]{#1}%
\providecommand \bibfnamefont [1]{#1}%
\providecommand \citenamefont [1]{#1}%
\providecommand \href@noop [0]{\@secondoftwo}%
\providecommand \href [0]{\begingroup \@sanitize@url \@href}%
\providecommand \@href[1]{\@@startlink{#1}\@@href}%
\providecommand \@@href[1]{\endgroup#1\@@endlink}%
\providecommand \@sanitize@url [0]{\catcode `\\12\catcode `\$12\catcode
  `\&12\catcode `\#12\catcode `\^12\catcode `\_12\catcode `\%12\relax}%
\providecommand \@@startlink[1]{}%
\providecommand \@@endlink[0]{}%
\providecommand \url  [0]{\begingroup\@sanitize@url \@url }%
\providecommand \@url [1]{\endgroup\@href {#1}{\urlprefix }}%
\providecommand \urlprefix  [0]{URL }%
\providecommand \Eprint [0]{\href }%
\providecommand \doibase [0]{http://dx.doi.org/}%
\providecommand \selectlanguage [0]{\@gobble}%
\providecommand \bibinfo  [0]{\@secondoftwo}%
\providecommand \bibfield  [0]{\@secondoftwo}%
\providecommand \translation [1]{[#1]}%
\providecommand \BibitemOpen [0]{}%
\providecommand \bibitemStop [0]{}%
\providecommand \bibitemNoStop [0]{.\EOS\space}%
\providecommand \EOS [0]{\spacefactor3000\relax}%
\providecommand \BibitemShut  [1]{\csname bibitem#1\endcsname}%
\let\auto@bib@innerbib\@empty
\bibitem [{\citenamefont {Meyer}(1976)}]{ref23}%
  \BibitemOpen
  \bibfield  {author} {\bibinfo {author} {\bibfnamefont {R.~B.}\ \bibnamefont
  {Meyer}},\ }\href@noop {} {\emph {\bibinfo {title} {Proceedings of the Les
  Houches Summer School on Theoretical Physics, 1973, session No. XXV}}}\
  (\bibinfo  {publisher} {New York: Gordon and Breach},\ \bibinfo {year}
  {1976})\BibitemShut {NoStop}%
\bibitem [{\citenamefont {Cestari}\ \emph {et~al.}(2011)\citenamefont
  {Cestari}, \citenamefont {Diez-Berart}, \citenamefont {Dunmur}, \citenamefont
  {Ferrarini}, \citenamefont {de~{La Fuente}}, \citenamefont {Jackson},
  \citenamefont {L\'opez}, \citenamefont {Luckhurst}, \citenamefont
  {Perez-Jubindo}, \citenamefont {Richardson}, \citenamefont {Salud},
  \citenamefont {Timimi},\ and\ \citenamefont
  {Zimmermann}}]{Cestari&DiezBerart2011}%
  \BibitemOpen
  \bibfield  {author} {\bibinfo {author} {\bibfnamefont {M.}~\bibnamefont
  {Cestari}}, \bibinfo {author} {\bibfnamefont {S.}~\bibnamefont
  {Diez-Berart}}, \bibinfo {author} {\bibfnamefont {D.~A.}\ \bibnamefont
  {Dunmur}}, \bibinfo {author} {\bibfnamefont {A.}~\bibnamefont {Ferrarini}},
  \bibinfo {author} {\bibfnamefont {M.~R.}\ \bibnamefont {de~{La Fuente}}},
  \bibinfo {author} {\bibfnamefont {D.~J.~B.}\ \bibnamefont {Jackson}},
  \bibinfo {author} {\bibfnamefont {D.~O.}\ \bibnamefont {L\'opez}}, \bibinfo
  {author} {\bibfnamefont {G.~R.}\ \bibnamefont {Luckhurst}}, \bibinfo {author}
  {\bibfnamefont {M.~A.}\ \bibnamefont {Perez-Jubindo}}, \bibinfo {author}
  {\bibfnamefont {R.~M.}\ \bibnamefont {Richardson}}, \bibinfo {author}
  {\bibfnamefont {J.}~\bibnamefont {Salud}}, \bibinfo {author} {\bibfnamefont
  {B.~A.}\ \bibnamefont {Timimi}}, \ and\ \bibinfo {author} {\bibfnamefont
  {H.}~\bibnamefont {Zimmermann}},\ }\href {\doibase
  10.1103/PhysRevE.84.031704} {\bibfield  {journal} {\bibinfo  {journal} {Phys.
  Rev. E}\ }\textbf {\bibinfo {volume} {84}},\ \bibinfo {pages} {031704}
  (\bibinfo {year} {2011})}\BibitemShut {NoStop}%
\bibitem [{\citenamefont {Chen}\ \emph
  {et~al.}(2013{\natexlab{a}})\citenamefont {Chen}, \citenamefont {Porada},
  \citenamefont {Hooper} \emph {et~al.}}]{ref2}%
  \BibitemOpen
  \bibfield  {author} {\bibinfo {author} {\bibfnamefont {D.}~\bibnamefont
  {Chen}}, \bibinfo {author} {\bibfnamefont {J.~H.}\ \bibnamefont {Porada}},
  \bibinfo {author} {\bibfnamefont {J.~B.}\ \bibnamefont {Hooper}},  \emph
  {et~al.},\ }\href {\doibase 10.1073/pnas.1314654110} {\bibfield  {journal}
  {\bibinfo  {journal} {Proc. Natl. Acad. Sci. USA}\ }\textbf {\bibinfo
  {volume} {110}},\ \bibinfo {pages} {15931} (\bibinfo {year}
  {2013}{\natexlab{a}})}\BibitemShut {NoStop}%
\bibitem [{\citenamefont {Borshch}\ \emph
  {et~al.}(2013{\natexlab{a}})\citenamefont {Borshch}, \citenamefont {Kim},
  \citenamefont {Xiang} \emph {et~al.}}]{ref1}%
  \BibitemOpen
  \bibfield  {author} {\bibinfo {author} {\bibfnamefont {V.}~\bibnamefont
  {Borshch}}, \bibinfo {author} {\bibfnamefont {Y.-K.}\ \bibnamefont {Kim}},
  \bibinfo {author} {\bibfnamefont {J.}~\bibnamefont {Xiang}},  \emph
  {et~al.},\ }\href {\doibase 10.1038/ncomms3635} {\bibfield  {journal}
  {\bibinfo  {journal} {Nat. Commun.}\ }\textbf {\bibinfo {volume} {4}},\
  \bibinfo {pages} {1} (\bibinfo {year} {2013}{\natexlab{a}})}\BibitemShut
  {NoStop}%
\bibitem [{\citenamefont {Mandle}(2016)}]{rewiew}%
  \BibitemOpen
  \bibfield  {author} {\bibinfo {author} {\bibfnamefont {R.~J.}\ \bibnamefont
  {Mandle}},\ }\href {\doibase 10.1039/C6SM01772J} {\bibfield  {journal}
  {\bibinfo  {journal} {Soft Matter}\ }\textbf {\bibinfo {volume} {12}},\
  \bibinfo {pages} {7883} (\bibinfo {year} {2016})}\BibitemShut {NoStop}%
\bibitem [{\citenamefont {Tavarone}, \citenamefont {Charbonneau},\ and\
  \citenamefont {Stark}(2015)}]{tavarone}%
  \BibitemOpen
  \bibfield  {author} {\bibinfo {author} {\bibfnamefont {R.}~\bibnamefont
  {Tavarone}}, \bibinfo {author} {\bibfnamefont {P.}~\bibnamefont
  {Charbonneau}}, \ and\ \bibinfo {author} {\bibfnamefont {H.}~\bibnamefont
  {Stark}},\ }\href {\doibase 10.1063/1.4930886} {\bibfield  {journal}
  {\bibinfo  {journal} {The Journal of Chemical Physics}\ }\textbf {\bibinfo
  {volume} {143}},\ \bibinfo {pages} {114505} (\bibinfo {year} {2015})},\
  \Eprint {http://arxiv.org/abs/https://doi.org/10.1063/1.4930886}
  {https://doi.org/10.1063/1.4930886} \BibitemShut {NoStop}%
\bibitem [{\citenamefont {Karbowniczek}\ \emph {et~al.}(2017)\citenamefont
  {Karbowniczek}, \citenamefont {Ciesla}, \citenamefont {Longa},\ and\
  \citenamefont {Chrzanowska}}]{karbowniczek}%
  \BibitemOpen
  \bibfield  {author} {\bibinfo {author} {\bibfnamefont {P.}~\bibnamefont
  {Karbowniczek}}, \bibinfo {author} {\bibfnamefont {M.}~\bibnamefont
  {Ciesla}}, \bibinfo {author} {\bibfnamefont {L.}~\bibnamefont {Longa}}, \
  and\ \bibinfo {author} {\bibfnamefont {A.}~\bibnamefont {Chrzanowska}},\
  }\href {\doibase 10.1080/02678292.2016.1259510} {\bibfield  {journal}
  {\bibinfo  {journal} {Liquid Crystals}\ }\textbf {\bibinfo {volume} {44}},\
  \bibinfo {pages} {254} (\bibinfo {year} {2017})},\ \Eprint
  {http://arxiv.org/abs/https://doi.org/10.1080/02678292.2016.1259510}
  {https://doi.org/10.1080/02678292.2016.1259510} \BibitemShut {NoStop}%
\bibitem [{\citenamefont {Karbowniczek}(2018)}]{karbowniczek2}%
  \BibitemOpen
  \bibfield  {author} {\bibinfo {author} {\bibfnamefont {P.}~\bibnamefont
  {Karbowniczek}},\ }\href {\doibase 10.1063/1.5021541} {\bibfield  {journal}
  {\bibinfo  {journal} {The Journal of Chemical Physics}\ }\textbf {\bibinfo
  {volume} {148}},\ \bibinfo {pages} {136101} (\bibinfo {year} {2018})},\
  \Eprint {http://arxiv.org/abs/https://doi.org/10.1063/1.5021541}
  {https://doi.org/10.1063/1.5021541} \BibitemShut {NoStop}%
\bibitem [{\citenamefont {Dozov}(2001)}]{ref25}%
  \BibitemOpen
  \bibfield  {author} {\bibinfo {author} {\bibfnamefont {I.}~\bibnamefont
  {Dozov}},\ }\href {\doibase 10.1209/epl/i2001-00513-x} {\bibfield  {journal}
  {\bibinfo  {journal} {Europhys. Lett.}\ }\textbf {\bibinfo {volume} {56}},\
  \bibinfo {pages} {247} (\bibinfo {year} {2001})}\BibitemShut {NoStop}%
\bibitem [{\citenamefont {de~Gennes}\ and\ \citenamefont
  {Prost}(1993)}]{ref36}%
  \BibitemOpen
  \bibfield  {author} {\bibinfo {author} {\bibfnamefont {P.~G.}\ \bibnamefont
  {de~Gennes}}\ and\ \bibinfo {author} {\bibfnamefont {J.}~\bibnamefont
  {Prost}},\ }\href@noop {} {\emph {\bibinfo {title} {The Physics of Liquid
  Crystals}}},\ \bibinfo {edition} {2nd}\ ed.\ (\bibinfo  {publisher}
  {Clarendon Press},\ \bibinfo {year} {1993})\BibitemShut {NoStop}%
\bibitem [{\citenamefont {J\'akli}, \citenamefont {Lavrentovich},\ and\
  \citenamefont {Selinger}(2018)}]{RevModPhys.90.045004}%
  \BibitemOpen
  \bibfield  {author} {\bibinfo {author} {\bibfnamefont {A.}~\bibnamefont
  {J\'akli}}, \bibinfo {author} {\bibfnamefont {O.~D.}\ \bibnamefont
  {Lavrentovich}}, \ and\ \bibinfo {author} {\bibfnamefont {J.~V.}\
  \bibnamefont {Selinger}},\ }\href {\doibase 10.1103/RevModPhys.90.045004}
  {\bibfield  {journal} {\bibinfo  {journal} {Rev. Mod. Phys.}\ }\textbf
  {\bibinfo {volume} {90}},\ \bibinfo {pages} {045004} (\bibinfo {year}
  {2018})}\BibitemShut {NoStop}%
\bibitem [{\citenamefont {Longa}\ and\ \citenamefont {Trebin}(1990)}]{ref48}%
  \BibitemOpen
  \bibfield  {author} {\bibinfo {author} {\bibfnamefont {L.}~\bibnamefont
  {Longa}}\ and\ \bibinfo {author} {\bibfnamefont {H.-R.}\ \bibnamefont
  {Trebin}},\ }\href {\doibase 10.1103/PhysRevA.42.3453} {\bibfield  {journal}
  {\bibinfo  {journal} {Phys. Rev. A}\ }\textbf {\bibinfo {volume} {42}},\
  \bibinfo {pages} {3453} (\bibinfo {year} {1990})}\BibitemShut {NoStop}%
\bibitem [{\citenamefont {Longa}\ and\ \citenamefont
  {Paj\c{a}k}(2016)}]{ref22}%
  \BibitemOpen
  \bibfield  {author} {\bibinfo {author} {\bibfnamefont {L.}~\bibnamefont
  {Longa}}\ and\ \bibinfo {author} {\bibfnamefont {G.}~\bibnamefont
  {Paj\c{a}k}},\ }\href {\doibase 10.1103/PhysRevE.93.040701} {\bibfield
  {journal} {\bibinfo  {journal} {Phys. Rev. E}\ }\textbf {\bibinfo {volume}
  {93}},\ \bibinfo {pages} {040701} (\bibinfo {year} {2016})}\BibitemShut
  {NoStop}%
\bibitem [{\citenamefont {Paj\c{a}k}, \citenamefont {Longa},\ and\
  \citenamefont {Chrzanowska}(2018)}]{extfield}%
  \BibitemOpen
  \bibfield  {author} {\bibinfo {author} {\bibfnamefont {G.}~\bibnamefont
  {Paj\c{a}k}}, \bibinfo {author} {\bibfnamefont {L.}~\bibnamefont {Longa}}, \
  and\ \bibinfo {author} {\bibfnamefont {A.}~\bibnamefont {Chrzanowska}},\
  }\href {\doibase 10.1073/pnas.1721786115} {\bibfield  {journal} {\bibinfo
  {journal} {Proceedings of the National Academy of Sciences}\ }\textbf
  {\bibinfo {volume} {115}},\ \bibinfo {pages} {E10303} (\bibinfo {year}
  {2018})},\ \Eprint
  {http://arxiv.org/abs/https://www.pnas.org/doi/pdf/10.1073/pnas.1721786115}
  {https://www.pnas.org/doi/pdf/10.1073/pnas.1721786115} \BibitemShut {NoStop}%
\bibitem [{\citenamefont {\v{S}epelj}\ \emph {et~al.}(2007)\citenamefont
  {\v{S}epelj}, \citenamefont {Lesac}, \citenamefont {Baumeister},
  \citenamefont {Diele}, \citenamefont {Nguyen},\ and\ \citenamefont
  {Bruce}}]{Sepelj&Lesac2007}%
  \BibitemOpen
  \bibfield  {author} {\bibinfo {author} {\bibfnamefont {M.}~\bibnamefont
  {\v{S}epelj}}, \bibinfo {author} {\bibfnamefont {A.}~\bibnamefont {Lesac}},
  \bibinfo {author} {\bibfnamefont {U.}~\bibnamefont {Baumeister}}, \bibinfo
  {author} {\bibfnamefont {S.}~\bibnamefont {Diele}}, \bibinfo {author}
  {\bibfnamefont {H.~L.}\ \bibnamefont {Nguyen}}, \ and\ \bibinfo {author}
  {\bibfnamefont {D.~W.}\ \bibnamefont {Bruce}},\ }\href {\doibase
  10.1039/B612517D} {\bibfield  {journal} {\bibinfo  {journal} {J. Mater.
  Chem.}\ }\textbf {\bibinfo {volume} {17}},\ \bibinfo {pages} {1154} (\bibinfo
  {year} {2007})}\BibitemShut {NoStop}%
\bibitem [{\citenamefont {Panov}\ \emph {et~al.}(2010)\citenamefont {Panov},
  \citenamefont {Nagaraj}, \citenamefont {Vij}, \citenamefont {Panarin},
  \citenamefont {Kohlmeier}, \citenamefont {Tamba}, \citenamefont {Lewis},\
  and\ \citenamefont {Mehl}}]{Panov&Nagaraj2010}%
  \BibitemOpen
  \bibfield  {author} {\bibinfo {author} {\bibfnamefont {V.~P.}\ \bibnamefont
  {Panov}}, \bibinfo {author} {\bibfnamefont {M.}~\bibnamefont {Nagaraj}},
  \bibinfo {author} {\bibfnamefont {J.~K.}\ \bibnamefont {Vij}}, \bibinfo
  {author} {\bibfnamefont {Y.~P.}\ \bibnamefont {Panarin}}, \bibinfo {author}
  {\bibfnamefont {A.}~\bibnamefont {Kohlmeier}}, \bibinfo {author}
  {\bibfnamefont {M.~G.}\ \bibnamefont {Tamba}}, \bibinfo {author}
  {\bibfnamefont {R.~A.}\ \bibnamefont {Lewis}}, \ and\ \bibinfo {author}
  {\bibfnamefont {G.~H.}\ \bibnamefont {Mehl}},\ }\href {\doibase
  10.1103/PhysRevLett.105.167801} {\bibfield  {journal} {\bibinfo  {journal}
  {Phys. Rev. Lett.}\ }\textbf {\bibinfo {volume} {105}},\ \bibinfo {pages}
  {167801} (\bibinfo {year} {2010})}\BibitemShut {NoStop}%
\bibitem [{\citenamefont {Borshch}\ \emph
  {et~al.}(2013{\natexlab{b}})\citenamefont {Borshch}, \citenamefont {Kim},
  \citenamefont {Xiang}, \citenamefont {Gao}, \citenamefont {J\'akli},
  \citenamefont {Panov}, \citenamefont {Vij}, \citenamefont {Imrie},
  \citenamefont {Tamba}, \citenamefont {Mehl},\ and\ \citenamefont
  {Lavrentovich}}]{Borshch&Kim2013}%
  \BibitemOpen
  \bibfield  {author} {\bibinfo {author} {\bibfnamefont {V.}~\bibnamefont
  {Borshch}}, \bibinfo {author} {\bibfnamefont {Y.-K.}\ \bibnamefont {Kim}},
  \bibinfo {author} {\bibfnamefont {J.}~\bibnamefont {Xiang}}, \bibinfo
  {author} {\bibfnamefont {M.}~\bibnamefont {Gao}}, \bibinfo {author}
  {\bibfnamefont {A.}~\bibnamefont {J\'akli}}, \bibinfo {author} {\bibfnamefont
  {V.~P.}\ \bibnamefont {Panov}}, \bibinfo {author} {\bibfnamefont {J.~K.}\
  \bibnamefont {Vij}}, \bibinfo {author} {\bibfnamefont {C.~T.}\ \bibnamefont
  {Imrie}}, \bibinfo {author} {\bibfnamefont {M.~G.}\ \bibnamefont {Tamba}},
  \bibinfo {author} {\bibfnamefont {G.~H.}\ \bibnamefont {Mehl}}, \ and\
  \bibinfo {author} {\bibfnamefont {O.~D.}\ \bibnamefont {Lavrentovich}},\
  }\href {\doibase 10.1038/ncomms3635} {\bibfield  {journal} {\bibinfo
  {journal} {Nat. Commun.}\ }\textbf {\bibinfo {volume} {4}},\ \bibinfo {pages}
  {2635} (\bibinfo {year} {2013}{\natexlab{b}})}\BibitemShut {NoStop}%
\bibitem [{\citenamefont {Chen}\ \emph
  {et~al.}(2013{\natexlab{b}})\citenamefont {Chen}, \citenamefont {Porada},
  \citenamefont {Hooper}, \citenamefont {Klittnick}, \citenamefont {Shen},
  \citenamefont {Tuchband}, \citenamefont {Korblova}, \citenamefont {Bedrov},
  \citenamefont {Walba}, \citenamefont {Glaser}, \citenamefont {Maclennan},\
  and\ \citenamefont {Clark}}]{Chen&Porada2013}%
  \BibitemOpen
  \bibfield  {author} {\bibinfo {author} {\bibfnamefont {D.}~\bibnamefont
  {Chen}}, \bibinfo {author} {\bibfnamefont {J.~H.}\ \bibnamefont {Porada}},
  \bibinfo {author} {\bibfnamefont {J.~B.}\ \bibnamefont {Hooper}}, \bibinfo
  {author} {\bibfnamefont {A.}~\bibnamefont {Klittnick}}, \bibinfo {author}
  {\bibfnamefont {Y.}~\bibnamefont {Shen}}, \bibinfo {author} {\bibfnamefont
  {M.~R.}\ \bibnamefont {Tuchband}}, \bibinfo {author} {\bibfnamefont
  {E.}~\bibnamefont {Korblova}}, \bibinfo {author} {\bibfnamefont
  {D.}~\bibnamefont {Bedrov}}, \bibinfo {author} {\bibfnamefont {D.~M.}\
  \bibnamefont {Walba}}, \bibinfo {author} {\bibfnamefont {M.~A.}\ \bibnamefont
  {Glaser}}, \bibinfo {author} {\bibfnamefont {J.~E.}\ \bibnamefont
  {Maclennan}}, \ and\ \bibinfo {author} {\bibfnamefont {N.~A.}\ \bibnamefont
  {Clark}},\ }\href {\doibase 10.1073/pnas.1314654110} {\bibfield  {journal}
  {\bibinfo  {journal} {Proc. Natl. Acad. Sci. USA}\ }\textbf {\bibinfo
  {volume} {110}},\ \bibinfo {pages} {15931} (\bibinfo {year}
  {2013}{\natexlab{b}})}\BibitemShut {NoStop}%
\bibitem [{\citenamefont {Paterson}\ \emph {et~al.}(2016)\citenamefont
  {Paterson}, \citenamefont {Gao}, \citenamefont {Kim}, \citenamefont {Jamali},
  \citenamefont {Finley}, \citenamefont {Robles-Hern\'andez}, \citenamefont
  {Diez-Berart}, \citenamefont {Salud}, \citenamefont {de~{La Fuente}},
  \citenamefont {Timimi}, \citenamefont {Zimmermann}, \citenamefont {Greco},
  \citenamefont {Ferrarini}, \citenamefont {Storey}, \citenamefont {L\'opez},
  \citenamefont {Lavrentovich}, \citenamefont {Luckhurst},\ and\ \citenamefont
  {Imrie}}]{Paterson&Gao2016}%
  \BibitemOpen
  \bibfield  {author} {\bibinfo {author} {\bibfnamefont {D.~A.}\ \bibnamefont
  {Paterson}}, \bibinfo {author} {\bibfnamefont {M.}~\bibnamefont {Gao}},
  \bibinfo {author} {\bibfnamefont {Y.-K.}\ \bibnamefont {Kim}}, \bibinfo
  {author} {\bibfnamefont {A.}~\bibnamefont {Jamali}}, \bibinfo {author}
  {\bibfnamefont {K.~L.}\ \bibnamefont {Finley}}, \bibinfo {author}
  {\bibfnamefont {B.}~\bibnamefont {Robles-Hern\'andez}}, \bibinfo {author}
  {\bibfnamefont {S.}~\bibnamefont {Diez-Berart}}, \bibinfo {author}
  {\bibfnamefont {J.}~\bibnamefont {Salud}}, \bibinfo {author} {\bibfnamefont
  {M.~R.}\ \bibnamefont {de~{La Fuente}}}, \bibinfo {author} {\bibfnamefont
  {B.~A.}\ \bibnamefont {Timimi}}, \bibinfo {author} {\bibfnamefont
  {H.}~\bibnamefont {Zimmermann}}, \bibinfo {author} {\bibfnamefont
  {C.}~\bibnamefont {Greco}}, \bibinfo {author} {\bibfnamefont
  {A.}~\bibnamefont {Ferrarini}}, \bibinfo {author} {\bibfnamefont {J.~M.~D.}\
  \bibnamefont {Storey}}, \bibinfo {author} {\bibfnamefont {D.~O.}\
  \bibnamefont {L\'opez}}, \bibinfo {author} {\bibfnamefont {O.~D.}\
  \bibnamefont {Lavrentovich}}, \bibinfo {author} {\bibfnamefont {G.~R.}\
  \bibnamefont {Luckhurst}}, \ and\ \bibinfo {author} {\bibfnamefont {C.~T.}\
  \bibnamefont {Imrie}},\ }\href {\doibase 10.1039/C6SM00537C} {\bibfield
  {journal} {\bibinfo  {journal} {Soft Matter}\ }\textbf {\bibinfo {volume}
  {12}},\ \bibinfo {pages} {6827} (\bibinfo {year} {2016})}\BibitemShut
  {NoStop}%
\bibitem [{\citenamefont {L\'opez}\ \emph {et~al.}(2016)\citenamefont
  {L\'opez}, \citenamefont {Robles-Hern\'andez}, \citenamefont {Salud},
  \citenamefont {de~{La Fuente}}, \citenamefont {Sebastian}, \citenamefont
  {Diez-Berart}, \citenamefont {Jaen}, \citenamefont {Dunmur},\ and\
  \citenamefont {Luckhurst}}]{Lopez&RoblesHernandez2016}%
  \BibitemOpen
  \bibfield  {author} {\bibinfo {author} {\bibfnamefont {D.~O.}\ \bibnamefont
  {L\'opez}}, \bibinfo {author} {\bibfnamefont {B.}~\bibnamefont
  {Robles-Hern\'andez}}, \bibinfo {author} {\bibfnamefont {J.}~\bibnamefont
  {Salud}}, \bibinfo {author} {\bibfnamefont {M.~R.}\ \bibnamefont {de~{La
  Fuente}}}, \bibinfo {author} {\bibfnamefont {N.}~\bibnamefont {Sebastian}},
  \bibinfo {author} {\bibfnamefont {S.}~\bibnamefont {Diez-Berart}}, \bibinfo
  {author} {\bibfnamefont {X.}~\bibnamefont {Jaen}}, \bibinfo {author}
  {\bibfnamefont {D.~A.}\ \bibnamefont {Dunmur}}, \ and\ \bibinfo {author}
  {\bibfnamefont {G.~R.}\ \bibnamefont {Luckhurst}},\ }\href {\doibase
  10.1039/C5CP07605F} {\bibfield  {journal} {\bibinfo  {journal} {Phys. Chem.
  Chem. Phys.}\ }\textbf {\bibinfo {volume} {18}},\ \bibinfo {pages} {4394}
  (\bibinfo {year} {2016})}\BibitemShut {NoStop}%
\bibitem [{\citenamefont {G{\"o}rtz}\ \emph {et~al.}(2009)\citenamefont
  {G{\"o}rtz}, \citenamefont {Southern}, \citenamefont {Roberts}, \citenamefont
  {Gleeson},\ and\ \citenamefont {Goodby}}]{Gortz&Southern2009}%
  \BibitemOpen
  \bibfield  {author} {\bibinfo {author} {\bibfnamefont {V.}~\bibnamefont
  {G{\"o}rtz}}, \bibinfo {author} {\bibfnamefont {C.}~\bibnamefont {Southern}},
  \bibinfo {author} {\bibfnamefont {N.~W.}\ \bibnamefont {Roberts}}, \bibinfo
  {author} {\bibfnamefont {H.~F.}\ \bibnamefont {Gleeson}}, \ and\ \bibinfo
  {author} {\bibfnamefont {J.~W.}\ \bibnamefont {Goodby}},\ }\href {\doibase
  10.1039/B808283A} {\bibfield  {journal} {\bibinfo  {journal} {Soft Matter}\
  }\textbf {\bibinfo {volume} {5}},\ \bibinfo {pages} {463} (\bibinfo {year}
  {2009})}\BibitemShut {NoStop}%
\bibitem [{\citenamefont {Chen}\ \emph {et~al.}(2014)\citenamefont {Chen},
  \citenamefont {Nakata}, \citenamefont {Shao}, \citenamefont {Tuchband},
  \citenamefont {Shuai}, \citenamefont {Baumeister}, \citenamefont {Weissflog},
  \citenamefont {Walba}, \citenamefont {Glaser}, \citenamefont {Maclennan},\
  and\ \citenamefont {Clark}}]{Chen&Nakata2014}%
  \BibitemOpen
  \bibfield  {author} {\bibinfo {author} {\bibfnamefont {D.}~\bibnamefont
  {Chen}}, \bibinfo {author} {\bibfnamefont {M.}~\bibnamefont {Nakata}},
  \bibinfo {author} {\bibfnamefont {R.}~\bibnamefont {Shao}}, \bibinfo {author}
  {\bibfnamefont {M.~R.}\ \bibnamefont {Tuchband}}, \bibinfo {author}
  {\bibfnamefont {M.}~\bibnamefont {Shuai}}, \bibinfo {author} {\bibfnamefont
  {U.}~\bibnamefont {Baumeister}}, \bibinfo {author} {\bibfnamefont
  {W.}~\bibnamefont {Weissflog}}, \bibinfo {author} {\bibfnamefont {D.~M.}\
  \bibnamefont {Walba}}, \bibinfo {author} {\bibfnamefont {M.~A.}\ \bibnamefont
  {Glaser}}, \bibinfo {author} {\bibfnamefont {J.~E.}\ \bibnamefont
  {Maclennan}}, \ and\ \bibinfo {author} {\bibfnamefont {N.~A.}\ \bibnamefont
  {Clark}},\ }\href {\doibase 10.1103/PhysRevE.89.022506} {\bibfield  {journal}
  {\bibinfo  {journal} {Phys. Rev. E}\ }\textbf {\bibinfo {volume} {89}},\
  \bibinfo {pages} {022506} (\bibinfo {year} {2014})}\BibitemShut {NoStop}%
\bibitem [{\citenamefont {Wang}\ \emph {et~al.}(2015)\citenamefont {Wang},
  \citenamefont {Singh}, \citenamefont {Agra-Kooijman}, \citenamefont {Gao},
  \citenamefont {Bisoyi}, \citenamefont {Xue}, \citenamefont {Fisch},
  \citenamefont {Kumar},\ and\ \citenamefont {Li}}]{trimer1}%
  \BibitemOpen
  \bibfield  {author} {\bibinfo {author} {\bibfnamefont {Y.}~\bibnamefont
  {Wang}}, \bibinfo {author} {\bibfnamefont {G.}~\bibnamefont {Singh}},
  \bibinfo {author} {\bibfnamefont {D.~M.}\ \bibnamefont {Agra-Kooijman}},
  \bibinfo {author} {\bibfnamefont {M.}~\bibnamefont {Gao}}, \bibinfo {author}
  {\bibfnamefont {H.~K.}\ \bibnamefont {Bisoyi}}, \bibinfo {author}
  {\bibfnamefont {C.}~\bibnamefont {Xue}}, \bibinfo {author} {\bibfnamefont
  {M.~R.}\ \bibnamefont {Fisch}}, \bibinfo {author} {\bibfnamefont
  {S.}~\bibnamefont {Kumar}}, \ and\ \bibinfo {author} {\bibfnamefont
  {Q.}~\bibnamefont {Li}},\ }\href {\doibase 10.1039/C4CE02502D} {\bibfield
  {journal} {\bibinfo  {journal} {CrystEngComm}\ }\textbf {\bibinfo {volume}
  {17}},\ \bibinfo {pages} {2778} (\bibinfo {year} {2015})}\BibitemShut
  {NoStop}%
\bibitem [{\citenamefont {Al-Janabi}, \citenamefont {Mandle},\ and\
  \citenamefont {Goodby}(2017)}]{trimer2}%
  \BibitemOpen
  \bibfield  {author} {\bibinfo {author} {\bibfnamefont {A.}~\bibnamefont
  {Al-Janabi}}, \bibinfo {author} {\bibfnamefont {R.~J.}\ \bibnamefont
  {Mandle}}, \ and\ \bibinfo {author} {\bibfnamefont {J.}~\bibnamefont
  {Goodby}},\ }\href {\doibase 10.1039/C7RA10261E} {\bibfield  {journal}
  {\bibinfo  {journal} {RSC Adv.}\ }\textbf {\bibinfo {volume} {7}},\ \bibinfo
  {pages} {47235} (\bibinfo {year} {2017})}\BibitemShut {NoStop}%
\bibitem [{\citenamefont {Mandle}\ and\ \citenamefont
  {Goodby}(2016)}]{oligomer}%
  \BibitemOpen
  \bibfield  {author} {\bibinfo {author} {\bibfnamefont {R.~J.}\ \bibnamefont
  {Mandle}}\ and\ \bibinfo {author} {\bibfnamefont {J.~W.}\ \bibnamefont
  {Goodby}},\ }\href {\doibase 10.1002/cphc.201600038} {\bibfield  {journal}
  {\bibinfo  {journal} {ChemPhysChem}\ }\textbf {\bibinfo {volume} {17}},\
  \bibinfo {pages} {967} (\bibinfo {year} {2016})}\BibitemShut {NoStop}%
\bibitem [{\citenamefont {Jansze}\ \emph {et~al.}(2015)\citenamefont {Jansze},
  \citenamefont {Martinez-Felipe}, \citenamefont {Storey}, \citenamefont
  {Marcelis},\ and\ \citenamefont {Imrie}}]{hydrogen1}%
  \BibitemOpen
  \bibfield  {author} {\bibinfo {author} {\bibfnamefont {S.~M.}\ \bibnamefont
  {Jansze}}, \bibinfo {author} {\bibfnamefont {A.}~\bibnamefont
  {Martinez-Felipe}}, \bibinfo {author} {\bibfnamefont {J.~M.~D.}\ \bibnamefont
  {Storey}}, \bibinfo {author} {\bibfnamefont {A.~T.~M.}\ \bibnamefont
  {Marcelis}}, \ and\ \bibinfo {author} {\bibfnamefont {C.~T.}\ \bibnamefont
  {Imrie}},\ }\href {\doibase 10.1002/anie.201409738} {\bibfield  {journal}
  {\bibinfo  {journal} {Angewandte Chemie - International Edition}\ }\textbf
  {\bibinfo {volume} {54}},\ \bibinfo {pages} {643} (\bibinfo {year}
  {2015})}\BibitemShut {NoStop}%
\bibitem [{\citenamefont {Fernández-Rico}\ \emph {et~al.}(2020)\citenamefont
  {Fernández-Rico}, \citenamefont {Chiappini}, \citenamefont {Yanagishima},
  \citenamefont {de~Sousa}, \citenamefont {Aarts}, \citenamefont {Dijkstra},\
  and\ \citenamefont {Dullens}}]{doi:10.1126/science.abb4536}%
  \BibitemOpen
  \bibfield  {author} {\bibinfo {author} {\bibfnamefont {C.}~\bibnamefont
  {Fernández-Rico}}, \bibinfo {author} {\bibfnamefont {M.}~\bibnamefont
  {Chiappini}}, \bibinfo {author} {\bibfnamefont {T.}~\bibnamefont
  {Yanagishima}}, \bibinfo {author} {\bibfnamefont {H.}~\bibnamefont
  {de~Sousa}}, \bibinfo {author} {\bibfnamefont {D.~G. A.~L.}\ \bibnamefont
  {Aarts}}, \bibinfo {author} {\bibfnamefont {M.}~\bibnamefont {Dijkstra}}, \
  and\ \bibinfo {author} {\bibfnamefont {R.~P.~A.}\ \bibnamefont {Dullens}},\
  }\href {\doibase 10.1126/science.abb4536} {\bibfield  {journal} {\bibinfo
  {journal} {Science}\ }\textbf {\bibinfo {volume} {369}},\ \bibinfo {pages}
  {950} (\bibinfo {year} {2020})},\ \Eprint
  {http://arxiv.org/abs/https://www.science.org/doi/pdf/10.1126/science.abb4536}
  {https://www.science.org/doi/pdf/10.1126/science.abb4536} \BibitemShut
  {NoStop}%
\bibitem [{\citenamefont {Merkel}\ \emph {et~al.}(2018)\citenamefont {Merkel},
  \citenamefont {Kocot}, \citenamefont {Vij},\ and\ \citenamefont
  {Shanker}}]{PhysRevE.98.022704}%
  \BibitemOpen
  \bibfield  {author} {\bibinfo {author} {\bibfnamefont {K.}~\bibnamefont
  {Merkel}}, \bibinfo {author} {\bibfnamefont {A.}~\bibnamefont {Kocot}},
  \bibinfo {author} {\bibfnamefont {J.~K.}\ \bibnamefont {Vij}}, \ and\
  \bibinfo {author} {\bibfnamefont {G.}~\bibnamefont {Shanker}},\ }\href
  {\doibase 10.1103/PhysRevE.98.022704} {\bibfield  {journal} {\bibinfo
  {journal} {Phys. Rev. E}\ }\textbf {\bibinfo {volume} {98}},\ \bibinfo
  {pages} {022704} (\bibinfo {year} {2018})}\BibitemShut {NoStop}%
\bibitem [{\citenamefont {Meyer}\ \emph {et~al.}(2020)\citenamefont {Meyer},
  \citenamefont {Blanc}, \citenamefont {Luckhurst}, \citenamefont {Davidson},\
  and\ \citenamefont {Dozov}}]{doi:10.1126/sciadv.abb8212}%
  \BibitemOpen
  \bibfield  {author} {\bibinfo {author} {\bibfnamefont {C.}~\bibnamefont
  {Meyer}}, \bibinfo {author} {\bibfnamefont {C.}~\bibnamefont {Blanc}},
  \bibinfo {author} {\bibfnamefont {G.~R.}\ \bibnamefont {Luckhurst}}, \bibinfo
  {author} {\bibfnamefont {P.}~\bibnamefont {Davidson}}, \ and\ \bibinfo
  {author} {\bibfnamefont {I.}~\bibnamefont {Dozov}},\ }\href {\doibase
  10.1126/sciadv.abb8212} {\bibfield  {journal} {\bibinfo  {journal} {Science
  Advances}\ }\textbf {\bibinfo {volume} {6}},\ \bibinfo {pages} {eabb8212}
  (\bibinfo {year} {2020})},\ \Eprint
  {http://arxiv.org/abs/https://www.science.org/doi/pdf/10.1126/sciadv.abb8212}
  {https://www.science.org/doi/pdf/10.1126/sciadv.abb8212} \BibitemShut
  {NoStop}%
\bibitem [{\citenamefont {Meyer}, \citenamefont {Luckhurst},\ and\
  \citenamefont {Dozov}(2015)}]{ref10}%
  \BibitemOpen
  \bibfield  {author} {\bibinfo {author} {\bibfnamefont {C.}~\bibnamefont
  {Meyer}}, \bibinfo {author} {\bibfnamefont {G.~R.}\ \bibnamefont
  {Luckhurst}}, \ and\ \bibinfo {author} {\bibfnamefont {I.}~\bibnamefont
  {Dozov}},\ }\href {\doibase 10.1039/C4TC01927J} {\bibfield  {journal}
  {\bibinfo  {journal} {J. Mater. Chem. C}\ }\textbf {\bibinfo {volume} {3}},\
  \bibinfo {pages} {318} (\bibinfo {year} {2015})}\BibitemShut {NoStop}%
\bibitem [{\citenamefont {Yun}\ \emph {et~al.}(2015)\citenamefont {Yun},
  \citenamefont {Vengatesan}, \citenamefont {Vij},\ and\ \citenamefont
  {Song}}]{elastic1}%
  \BibitemOpen
  \bibfield  {author} {\bibinfo {author} {\bibfnamefont {C.-J.}\ \bibnamefont
  {Yun}}, \bibinfo {author} {\bibfnamefont {M.~R.}\ \bibnamefont {Vengatesan}},
  \bibinfo {author} {\bibfnamefont {J.~K.}\ \bibnamefont {Vij}}, \ and\
  \bibinfo {author} {\bibfnamefont {J.-K.}\ \bibnamefont {Song}},\ }\href
  {\doibase 10.1063/1.4919065} {\bibfield  {journal} {\bibinfo  {journal}
  {Applied Physics Letters}\ }\textbf {\bibinfo {volume} {106}},\ \bibinfo
  {pages} {173102} (\bibinfo {year} {2015})},\ \Eprint
  {http://arxiv.org/abs/https://doi.org/10.1063/1.4919065}
  {https://doi.org/10.1063/1.4919065} \BibitemShut {NoStop}%
\bibitem [{\citenamefont {Babakhanova}\ \emph {et~al.}(2017)\citenamefont
  {Babakhanova}, \citenamefont {Parsouzi}, \citenamefont {Paladugu},
  \citenamefont {Wang}, \citenamefont {Nastishin}, \citenamefont
  {Shiyanovskii}, \citenamefont {Sprunt},\ and\ \citenamefont
  {Lavrentovich}}]{elastic2}%
  \BibitemOpen
  \bibfield  {author} {\bibinfo {author} {\bibfnamefont {G.}~\bibnamefont
  {Babakhanova}}, \bibinfo {author} {\bibfnamefont {Z.}~\bibnamefont
  {Parsouzi}}, \bibinfo {author} {\bibfnamefont {S.}~\bibnamefont {Paladugu}},
  \bibinfo {author} {\bibfnamefont {H.}~\bibnamefont {Wang}}, \bibinfo {author}
  {\bibfnamefont {Y.~A.}\ \bibnamefont {Nastishin}}, \bibinfo {author}
  {\bibfnamefont {S.~V.}\ \bibnamefont {Shiyanovskii}}, \bibinfo {author}
  {\bibfnamefont {S.}~\bibnamefont {Sprunt}}, \ and\ \bibinfo {author}
  {\bibfnamefont {O.~D.}\ \bibnamefont {Lavrentovich}},\ }\href {\doibase
  10.1103/PhysRevE.96.062704} {\bibfield  {journal} {\bibinfo  {journal} {Phys.
  Rev. E}\ }\textbf {\bibinfo {volume} {96}},\ \bibinfo {pages} {062704}
  (\bibinfo {year} {2017})}\BibitemShut {NoStop}%
\bibitem [{\citenamefont {Archbold}\ \emph {et~al.}(2017)\citenamefont
  {Archbold}, \citenamefont {Mandle}, \citenamefont {Andrews}, \citenamefont
  {Cowling},\ and\ \citenamefont {Goodby}}]{Archbold2017}%
  \BibitemOpen
  \bibfield  {author} {\bibinfo {author} {\bibfnamefont {C.~T.}\ \bibnamefont
  {Archbold}}, \bibinfo {author} {\bibfnamefont {R.~J.}\ \bibnamefont
  {Mandle}}, \bibinfo {author} {\bibfnamefont {J.~L.}\ \bibnamefont {Andrews}},
  \bibinfo {author} {\bibfnamefont {S.~J.}\ \bibnamefont {Cowling}}, \ and\
  \bibinfo {author} {\bibfnamefont {J.~W.}\ \bibnamefont {Goodby}},\ }\href
  {\doibase 10.1080/02678292.2017.1360954} {\bibfield  {journal} {\bibinfo
  {journal} {Liquid Crystals}\ }\textbf {\bibinfo {volume} {00}},\ \bibinfo
  {pages} {1} (\bibinfo {year} {2017})}\BibitemShut {NoStop}%
\bibitem [{\citenamefont {J\'ozefowicz}\ and\ \citenamefont
  {Longa}(2007)}]{doi:10.1080/15421400701738586}%
  \BibitemOpen
  \bibfield  {author} {\bibinfo {author} {\bibfnamefont {W.}~\bibnamefont
  {J\'ozefowicz}}\ and\ \bibinfo {author} {\bibfnamefont {L.}~\bibnamefont
  {Longa}},\ }\href {\doibase 10.1080/15421400701738586} {\bibfield  {journal}
  {\bibinfo  {journal} {Molecular Crystals and Liquid Crystals}\ }\textbf
  {\bibinfo {volume} {478}},\ \bibinfo {pages} {115/[871]} (\bibinfo {year}
  {2007})},\ \Eprint
  {http://arxiv.org/abs/https://doi.org/10.1080/15421400701738586}
  {https://doi.org/10.1080/15421400701738586} \BibitemShut {NoStop}%
\bibitem [{\citenamefont {Vaupoti{\v{c}}}\ \emph {et~al.}(2014)\citenamefont
  {Vaupoti{\v{c}}}, \citenamefont {{\v{C}}epi{\v{c}}}, \citenamefont {Osipov},\
  and\ \citenamefont {Gorecka}}]{Vaupotic2014}%
  \BibitemOpen
  \bibfield  {author} {\bibinfo {author} {\bibfnamefont {N.}~\bibnamefont
  {Vaupoti{\v{c}}}}, \bibinfo {author} {\bibfnamefont {M.}~\bibnamefont
  {{\v{C}}epi{\v{c}}}}, \bibinfo {author} {\bibfnamefont {M.~A.}\ \bibnamefont
  {Osipov}}, \ and\ \bibinfo {author} {\bibfnamefont {E.}~\bibnamefont
  {Gorecka}},\ }\href {\doibase 10.1103/PhysRevE.89.030501} {\bibfield
  {journal} {\bibinfo  {journal} {Physical Review E - Statistical, Nonlinear,
  and Soft Matter Physics}\ }\textbf {\bibinfo {volume} {89}},\ \bibinfo
  {pages} {2} (\bibinfo {year} {2014})}\BibitemShut {NoStop}%
\bibitem [{\citenamefont {Lubensky}\ and\ \citenamefont
  {Radzihovsky}(2002)}]{ref35}%
  \BibitemOpen
  \bibfield  {author} {\bibinfo {author} {\bibfnamefont {T.~C.}\ \bibnamefont
  {Lubensky}}\ and\ \bibinfo {author} {\bibfnamefont {L.}~\bibnamefont
  {Radzihovsky}},\ }\href {\doibase 10.1103/PhysRevE.66.031704} {\bibfield
  {journal} {\bibinfo  {journal} {Phys. Rev. E}\ }\textbf {\bibinfo {volume}
  {66}},\ \bibinfo {pages} {031704} (\bibinfo {year} {2002})}\BibitemShut
  {NoStop}%
\bibitem [{\citenamefont {Goodby}(2017)}]{Goodby}%
  \BibitemOpen
  \bibfield  {author} {\bibinfo {author} {\bibfnamefont {J.~W.}\ \bibnamefont
  {Goodby}},\ }\href {\doibase 10.1080/02678292.2017.1347293} {\bibfield
  {journal} {\bibinfo  {journal} {Liquid Crystals}\ }\textbf {\bibinfo {volume}
  {44}},\ \bibinfo {pages} {1755} (\bibinfo {year} {2017})},\ \Eprint
  {http://arxiv.org/abs/https://doi.org/10.1080/02678292.2017.1347293}
  {https://doi.org/10.1080/02678292.2017.1347293} \BibitemShut {NoStop}%
\bibitem [{\citenamefont {Delaire}\ and\ \citenamefont
  {Nakatani}(2000)}]{doi:10.1021/cr980078m}%
  \BibitemOpen
  \bibfield  {author} {\bibinfo {author} {\bibfnamefont {J.~A.}\ \bibnamefont
  {Delaire}}\ and\ \bibinfo {author} {\bibfnamefont {K.}~\bibnamefont
  {Nakatani}},\ }\href {\doibase 10.1021/cr980078m} {\bibfield  {journal}
  {\bibinfo  {journal} {Chemical Reviews}\ }\textbf {\bibinfo {volume} {100}},\
  \bibinfo {pages} {1817} (\bibinfo {year} {2000})},\ \bibinfo {note} {pMID:
  11777422},\ \Eprint {http://arxiv.org/abs/https://doi.org/10.1021/cr980078m}
  {https://doi.org/10.1021/cr980078m} \BibitemShut {NoStop}%
\bibitem [{\citenamefont {Katsonis}\ \emph {et~al.}(2007)\citenamefont
  {Katsonis}, \citenamefont {Lubomska}, \citenamefont {Pollard}, \citenamefont
  {Feringa},\ and\ \citenamefont {Rudolf}}]{KATSONIS2007407}%
  \BibitemOpen
  \bibfield  {author} {\bibinfo {author} {\bibfnamefont {N.}~\bibnamefont
  {Katsonis}}, \bibinfo {author} {\bibfnamefont {M.}~\bibnamefont {Lubomska}},
  \bibinfo {author} {\bibfnamefont {M.~M.}\ \bibnamefont {Pollard}}, \bibinfo
  {author} {\bibfnamefont {B.~L.}\ \bibnamefont {Feringa}}, \ and\ \bibinfo
  {author} {\bibfnamefont {P.}~\bibnamefont {Rudolf}},\ }\href {\doibase
  https://doi.org/10.1016/j.progsurf.2007.03.011} {\bibfield  {journal}
  {\bibinfo  {journal} {Progress in Surface Science}\ }\textbf {\bibinfo
  {volume} {82}},\ \bibinfo {pages} {407} (\bibinfo {year} {2007})}\BibitemShut
  {NoStop}%
\bibitem [{\citenamefont {Browne}\ and\ \citenamefont
  {Feringa}(2009)}]{doi:10.1146/annurev.physchem.040808.090423}%
  \BibitemOpen
  \bibfield  {author} {\bibinfo {author} {\bibfnamefont {W.~R.}\ \bibnamefont
  {Browne}}\ and\ \bibinfo {author} {\bibfnamefont {B.~L.}\ \bibnamefont
  {Feringa}},\ }\href {\doibase 10.1146/annurev.physchem.040808.090423}
  {\bibfield  {journal} {\bibinfo  {journal} {Annual Review of Physical
  Chemistry}\ }\textbf {\bibinfo {volume} {60}},\ \bibinfo {pages} {407}
  (\bibinfo {year} {2009})},\ \bibinfo {note} {pMID: 18999995},\ \Eprint
  {http://arxiv.org/abs/https://doi.org/10.1146/annurev.physchem.040808.090423}
  {https://doi.org/10.1146/annurev.physchem.040808.090423} \BibitemShut
  {NoStop}%
\bibitem [{\citenamefont {Fang}\ \emph {et~al.}(2010)\citenamefont {Fang},
  \citenamefont {Shi}, \citenamefont {Maclennan}, \citenamefont {Clark},
  \citenamefont {Farrow},\ and\ \citenamefont {Walba}}]{doi:10.1021/la102788j}%
  \BibitemOpen
  \bibfield  {author} {\bibinfo {author} {\bibfnamefont {G.}~\bibnamefont
  {Fang}}, \bibinfo {author} {\bibfnamefont {Y.}~\bibnamefont {Shi}}, \bibinfo
  {author} {\bibfnamefont {J.~E.}\ \bibnamefont {Maclennan}}, \bibinfo {author}
  {\bibfnamefont {N.~A.}\ \bibnamefont {Clark}}, \bibinfo {author}
  {\bibfnamefont {M.~J.}\ \bibnamefont {Farrow}}, \ and\ \bibinfo {author}
  {\bibfnamefont {D.~M.}\ \bibnamefont {Walba}},\ }\href {\doibase
  10.1021/la102788j} {\bibfield  {journal} {\bibinfo  {journal} {Langmuir}\
  }\textbf {\bibinfo {volume} {26}},\ \bibinfo {pages} {17482} (\bibinfo {year}
  {2010})},\ \bibinfo {note} {pMID: 20929215},\ \Eprint
  {http://arxiv.org/abs/https://doi.org/10.1021/la102788j}
  {https://doi.org/10.1021/la102788j} \BibitemShut {NoStop}%
\bibitem [{\citenamefont {Wu}\ \emph {et~al.}(2021)\citenamefont {Wu},
  \citenamefont {Liu}, \citenamefont {Chen},\ and\ \citenamefont
  {Yang}}]{cryst11121560}%
  \BibitemOpen
  \bibfield  {author} {\bibinfo {author} {\bibfnamefont {Y.}~\bibnamefont
  {Wu}}, \bibinfo {author} {\bibfnamefont {Y.}~\bibnamefont {Liu}}, \bibinfo
  {author} {\bibfnamefont {J.}~\bibnamefont {Chen}}, \ and\ \bibinfo {author}
  {\bibfnamefont {R.}~\bibnamefont {Yang}},\ }\href {\doibase
  10.3390/cryst11121560} {\bibfield  {journal} {\bibinfo  {journal} {Crystals}\
  }\textbf {\bibinfo {volume} {11}} (\bibinfo {year} {2021}),\
  10.3390/cryst11121560}\BibitemShut {NoStop}%
\bibitem [{\citenamefont {Pe\'on}\ \emph {et~al.}(2006)\citenamefont {Pe\'on},
  \citenamefont {Saucedo-Zugazagoitia}, \citenamefont {Pucheta-Mendez},
  \citenamefont {Perusqu\'ia}, \citenamefont {Sutmann},\ and\ \citenamefont
  {Quintana-H}}]{peonek}%
  \BibitemOpen
  \bibfield  {author} {\bibinfo {author} {\bibfnamefont {J.}~\bibnamefont
  {Pe\'on}}, \bibinfo {author} {\bibfnamefont {J.}~\bibnamefont
  {Saucedo-Zugazagoitia}}, \bibinfo {author} {\bibfnamefont {F.}~\bibnamefont
  {Pucheta-Mendez}}, \bibinfo {author} {\bibfnamefont {R.~A.}\ \bibnamefont
  {Perusqu\'ia}}, \bibinfo {author} {\bibfnamefont {G.}~\bibnamefont
  {Sutmann}}, \ and\ \bibinfo {author} {\bibfnamefont {J.}~\bibnamefont
  {Quintana-H}},\ }\href {\doibase 10.1063/1.2338313} {\bibfield  {journal}
  {\bibinfo  {journal} {The Journal of Chemical Physics}\ }\textbf {\bibinfo
  {volume} {125}},\ \bibinfo {pages} {104908} (\bibinfo {year} {2006})},\
  \Eprint {http://arxiv.org/abs/https://doi.org/10.1063/1.2338313}
  {https://doi.org/10.1063/1.2338313} \BibitemShut {NoStop}%
\bibitem [{\citenamefont {Varga}\ \emph {et~al.}(2009)\citenamefont {Varga},
  \citenamefont {Gurin}, \citenamefont {Armas-P\'erez},\ and\ \citenamefont
  {Quintana-H}}]{varga}%
  \BibitemOpen
  \bibfield  {author} {\bibinfo {author} {\bibfnamefont {S.}~\bibnamefont
  {Varga}}, \bibinfo {author} {\bibfnamefont {P.}~\bibnamefont {Gurin}},
  \bibinfo {author} {\bibfnamefont {J.~C.}\ \bibnamefont {Armas-P\'erez}}, \
  and\ \bibinfo {author} {\bibfnamefont {J.}~\bibnamefont {Quintana-H}},\
  }\href {\doibase 10.1063/1.3258858} {\bibfield  {journal} {\bibinfo
  {journal} {The Journal of Chemical Physics}\ }\textbf {\bibinfo {volume}
  {131}},\ \bibinfo {pages} {184901} (\bibinfo {year} {2009})},\ \Eprint
  {http://arxiv.org/abs/https://doi.org/10.1063/1.3258858}
  {https://doi.org/10.1063/1.3258858} \BibitemShut {NoStop}%
\bibitem [{\citenamefont {J.~Mart\'inez-Gonz\'alez}(2012)}]{martinez}%
  \BibitemOpen
  \bibfield  {author} {\bibinfo {author} {\bibfnamefont {J.~Q.-H.}\
  \bibnamefont {J.~Mart\'inez-Gonz\'alez}, \bibfnamefont {J.~C.
  Armas-P\'erez}},\ }\href@noop {} {\bibfield  {journal} {\bibinfo  {journal}
  {J. Stat. Phys.}\ }\textbf {\bibinfo {volume} {150}},\ \bibinfo {pages} {559}
  (\bibinfo {year} {2012})}\BibitemShut {NoStop}%
\bibitem [{\citenamefont {J\'ozefowicz}\ and\ \citenamefont
  {Longa}(2011)}]{doi:10.1080/15421406.2011.572014}%
  \BibitemOpen
  \bibfield  {author} {\bibinfo {author} {\bibfnamefont {W.}~\bibnamefont
  {J\'ozefowicz}}\ and\ \bibinfo {author} {\bibfnamefont {L.}~\bibnamefont
  {Longa}},\ }\href {\doibase 10.1080/15421406.2011.572014} {\bibfield
  {journal} {\bibinfo  {journal} {Molecular Crystals and Liquid Crystals}\
  }\textbf {\bibinfo {volume} {545}},\ \bibinfo {pages} {204/[1428]} (\bibinfo
  {year} {2011})},\ \Eprint
  {http://arxiv.org/abs/https://doi.org/10.1080/15421406.2011.572014}
  {https://doi.org/10.1080/15421406.2011.572014} \BibitemShut {NoStop}%
\bibitem [{\citenamefont {Cinacchi}\ \emph {et~al.}(2017)\citenamefont
  {Cinacchi}, \citenamefont {Ferrarini}, \citenamefont {Giacometti},\ and\
  \citenamefont {Kolli}}]{cinacchi}%
  \BibitemOpen
  \bibfield  {author} {\bibinfo {author} {\bibfnamefont {G.}~\bibnamefont
  {Cinacchi}}, \bibinfo {author} {\bibfnamefont {A.}~\bibnamefont {Ferrarini}},
  \bibinfo {author} {\bibfnamefont {A.}~\bibnamefont {Giacometti}}, \ and\
  \bibinfo {author} {\bibfnamefont {H.~B.}\ \bibnamefont {Kolli}},\ }\href
  {\doibase 10.1063/1.4996610} {\bibfield  {journal} {\bibinfo  {journal} {The
  Journal of Chemical Physics}\ }\textbf {\bibinfo {volume} {147}},\ \bibinfo
  {pages} {224903} (\bibinfo {year} {2017})},\ \Eprint
  {http://arxiv.org/abs/https://doi.org/10.1063/1.4996610}
  {https://doi.org/10.1063/1.4996610} \BibitemShut {NoStop}%
\bibitem [{\citenamefont {Schlotthauer}\ \emph {et~al.}(2015)\citenamefont
  {Schlotthauer}, \citenamefont {Skutnik}, \citenamefont {Stieger},\ and\
  \citenamefont {Schoen}}]{schlotthauer}%
  \BibitemOpen
  \bibfield  {author} {\bibinfo {author} {\bibfnamefont {S.}~\bibnamefont
  {Schlotthauer}}, \bibinfo {author} {\bibfnamefont {R.~A.}\ \bibnamefont
  {Skutnik}}, \bibinfo {author} {\bibfnamefont {T.}~\bibnamefont {Stieger}}, \
  and\ \bibinfo {author} {\bibfnamefont {M.}~\bibnamefont {Schoen}},\ }\href
  {\doibase 10.1063/1.4920979} {\bibfield  {journal} {\bibinfo  {journal} {The
  Journal of Chemical Physics}\ }\textbf {\bibinfo {volume} {142}},\ \bibinfo
  {pages} {194704} (\bibinfo {year} {2015})},\ \Eprint
  {http://arxiv.org/abs/https://doi.org/10.1063/1.4920979}
  {https://doi.org/10.1063/1.4920979} \BibitemShut {NoStop}%
\bibitem [{\citenamefont {Anzivino}, \citenamefont {van Roij},\ and\
  \citenamefont {Dijkstra}(2020)}]{doi:10.1063/5.0008936}%
  \BibitemOpen
  \bibfield  {author} {\bibinfo {author} {\bibfnamefont {C.}~\bibnamefont
  {Anzivino}}, \bibinfo {author} {\bibfnamefont {R.}~\bibnamefont {van Roij}},
  \ and\ \bibinfo {author} {\bibfnamefont {M.}~\bibnamefont {Dijkstra}},\
  }\href {\doibase 10.1063/5.0008936} {\bibfield  {journal} {\bibinfo
  {journal} {The Journal of Chemical Physics}\ }\textbf {\bibinfo {volume}
  {152}},\ \bibinfo {pages} {224502} (\bibinfo {year} {2020})},\ \Eprint
  {http://arxiv.org/abs/https://doi.org/10.1063/5.0008936}
  {https://doi.org/10.1063/5.0008936} \BibitemShut {NoStop}%
\bibitem [{\citenamefont {Anzivino}, \citenamefont {van Roij},\ and\
  \citenamefont {Dijkstra}(2022)}]{PhysRevE.105.L022701}%
  \BibitemOpen
  \bibfield  {author} {\bibinfo {author} {\bibfnamefont {C.}~\bibnamefont
  {Anzivino}}, \bibinfo {author} {\bibfnamefont {R.}~\bibnamefont {van Roij}},
  \ and\ \bibinfo {author} {\bibfnamefont {M.}~\bibnamefont {Dijkstra}},\
  }\href {\doibase 10.1103/PhysRevE.105.L022701} {\bibfield  {journal}
  {\bibinfo  {journal} {Phys. Rev. E}\ }\textbf {\bibinfo {volume} {105}},\
  \bibinfo {pages} {L022701} (\bibinfo {year} {2022})}\BibitemShut {NoStop}%
\bibitem [{\citenamefont {Chrzanowska}(2016)}]{Agnesferro}%
  \BibitemOpen
  \bibfield  {author} {\bibinfo {author} {\bibfnamefont {A.}~\bibnamefont
  {Chrzanowska}},\ }\href {\doibase 10.1080/00150193.2016.1136730} {\bibfield
  {journal} {\bibinfo  {journal} {Ferroelectrics}\ }\textbf {\bibinfo {volume}
  {495}},\ \bibinfo {pages} {43} (\bibinfo {year} {2016})},\ \Eprint
  {http://arxiv.org/abs/https://doi.org/10.1080/00150193.2016.1136730}
  {https://doi.org/10.1080/00150193.2016.1136730} \BibitemShut {NoStop}%
\bibitem [{\citenamefont {Chrzanowska}(2017)}]{AgnesSmek}%
  \BibitemOpen
  \bibfield  {author} {\bibinfo {author} {\bibfnamefont {A.}~\bibnamefont
  {Chrzanowska}},\ }\href {\doibase 10.1103/PhysRevE.95.063316} {\bibfield
  {journal} {\bibinfo  {journal} {Phys. Rev. E}\ }\textbf {\bibinfo {volume}
  {95}},\ \bibinfo {pages} {063316} (\bibinfo {year} {2017})}\BibitemShut
  {NoStop}%
\bibitem [{\citenamefont {Kayser}\ and\ \citenamefont
  {Ravech\'e}(1978)}]{bif1_Kayser}%
  \BibitemOpen
  \bibfield  {author} {\bibinfo {author} {\bibfnamefont {R.~F.}\ \bibnamefont
  {Kayser}}\ and\ \bibinfo {author} {\bibfnamefont {H.~J.}\ \bibnamefont
  {Ravech\'e}},\ }\href {\doibase 10.1103/PhysRevA.17.2067} {\bibfield
  {journal} {\bibinfo  {journal} {Phys. Rev. A}\ }\textbf {\bibinfo {volume}
  {17}},\ \bibinfo {pages} {2067} (\bibinfo {year} {1978})}\BibitemShut
  {NoStop}%
\bibitem [{\citenamefont {Mulder}(1989)}]{bif2_Mulder}%
  \BibitemOpen
  \bibfield  {author} {\bibinfo {author} {\bibfnamefont {B.}~\bibnamefont
  {Mulder}},\ }\href {\doibase 10.1103/PhysRevA.39.360} {\bibfield  {journal}
  {\bibinfo  {journal} {Phys. Rev. A}\ }\textbf {\bibinfo {volume} {39}},\
  \bibinfo {pages} {360} (\bibinfo {year} {1989})}\BibitemShut {NoStop}%
\bibitem [{\citenamefont {Longa}(1986{\natexlab{a}})}]{longa_bif1}%
  \BibitemOpen
  \bibfield  {author} {\bibinfo {author} {\bibfnamefont {L.}~\bibnamefont
  {Longa}},\ }\href {\doibase 10.1063/1.451007} {\bibfield  {journal} {\bibinfo
   {journal} {The Journal of Chemical Physics}\ }\textbf {\bibinfo {volume}
  {85}},\ \bibinfo {pages} {2974} (\bibinfo {year} {1986}{\natexlab{a}})},\
  \Eprint {http://arxiv.org/abs/https://doi.org/10.1063/1.451007}
  {https://doi.org/10.1063/1.451007} \BibitemShut {NoStop}%
\bibitem [{\citenamefont {Longa}(1986{\natexlab{b}})}]{longa_bif3}%
  \BibitemOpen
  \bibfield  {author} {\bibinfo {author} {\bibfnamefont {L.}~\bibnamefont
  {Longa}},\ }\href {\doibase 10.1007/BF01303606} {\bibfield  {journal}
  {\bibinfo  {journal} {Zeitschrift für Physik B Condensed Matter}\ }\textbf
  {\bibinfo {volume} {64}},\ \bibinfo {pages} {357} (\bibinfo {year}
  {1986}{\natexlab{b}})}\BibitemShut {NoStop}%
\bibitem [{\citenamefont {Longa}(1989)}]{longa_bif4}%
  \BibitemOpen
  \bibfield  {author} {\bibinfo {author} {\bibfnamefont {L.}~\bibnamefont
  {Longa}},\ }\href {\doibase 10.1080/02678298908045395} {\bibfield  {journal}
  {\bibinfo  {journal} {Liquid Crystals}\ }\textbf {\bibinfo {volume} {5}},\
  \bibinfo {pages} {443} (\bibinfo {year} {1989})},\ \Eprint
  {http://arxiv.org/abs/https://doi.org/10.1080/02678298908045395}
  {https://doi.org/10.1080/02678298908045395} \BibitemShut {NoStop}%
\bibitem [{\citenamefont {Longa}\ \emph {et~al.}(2005)\citenamefont {Longa},
  \citenamefont {Grzybowski}, \citenamefont {Romano},\ and\ \citenamefont
  {Virga}}]{longa_bif2}%
  \BibitemOpen
  \bibfield  {author} {\bibinfo {author} {\bibfnamefont {L.}~\bibnamefont
  {Longa}}, \bibinfo {author} {\bibfnamefont {P.}~\bibnamefont {Grzybowski}},
  \bibinfo {author} {\bibfnamefont {S.}~\bibnamefont {Romano}}, \ and\ \bibinfo
  {author} {\bibfnamefont {E.}~\bibnamefont {Virga}},\ }\href {\doibase
  10.1103/PhysRevE.71.051714} {\bibfield  {journal} {\bibinfo  {journal} {Phys.
  Rev. E}\ }\textbf {\bibinfo {volume} {71}},\ \bibinfo {pages} {051714}
  (\bibinfo {year} {2005})}\BibitemShut {NoStop}%
\bibitem [{\citenamefont {Onsager}(1949)}]{onsageraartykul}%
  \BibitemOpen
  \bibfield  {author} {\bibinfo {author} {\bibfnamefont {L.}~\bibnamefont
  {Onsager}},\ }\href {\doibase 10.1111/j.1749-6632.1949.tb27296.x} {\bibfield
  {journal} {\bibinfo  {journal} {Annals of the New York Academy of Sciences}\
  }\textbf {\bibinfo {volume} {51}},\ \bibinfo {pages} {627} (\bibinfo {year}
  {1949})}\BibitemShut {NoStop}%
\bibitem [{\citenamefont {Tavarone}, \citenamefont {Charbonneau},\ and\
  \citenamefont {Stark}(2016)}]{tavarone2}%
  \BibitemOpen
  \bibfield  {author} {\bibinfo {author} {\bibfnamefont {R.}~\bibnamefont
  {Tavarone}}, \bibinfo {author} {\bibfnamefont {P.}~\bibnamefont
  {Charbonneau}}, \ and\ \bibinfo {author} {\bibfnamefont {H.}~\bibnamefont
  {Stark}},\ }\href {\doibase 10.1063/1.4943393} {\bibfield  {journal}
  {\bibinfo  {journal} {The Journal of Chemical Physics}\ }\textbf {\bibinfo
  {volume} {144}},\ \bibinfo {pages} {104703} (\bibinfo {year} {2016})},\
  \Eprint {http://arxiv.org/abs/https://doi.org/10.1063/1.4943393}
  {https://doi.org/10.1063/1.4943393} \BibitemShut {NoStop}%
\end{thebibliography}%

\end{document}